\documentclass[10pt,prd,aps,nofootinbib,eqsecnum,notitlepage,nobibnotes,superscriptaddress,preprintnumbers,twocolumn]{revtex4-1}
\usepackage{hyperref}

\usepackage[a4paper, left=20mm,right=20mm,top=20mm,bottom=20mm]{geometry}
\usepackage[titletoc,toc,title]{appendix}
\usepackage{soul}

\usepackage{mathtools}
\usepackage{tensor}
\usepackage{derivative}
\usepackage{bm}
\usepackage{tabularray}
\usepackage{suffix}

\usepackage{amsmath, amsfonts,bm}
\usepackage{graphicx}
\usepackage{xcolor}

\DeclareMathAlphabet{\mathup}{OT1}{\familydefault}{m}{n}

\renewcommand\vec[1]{\bm{#1}}

\newcommand{\be}{\begin{equation}} 
\newcommand{\ee}{\end{equation}}

\begin{document}

\title{Close Hyperbolic Encounters in \textit{f(R)} Gravity}

\author{Jenna Bruton}
    \email{brtjen003@myuct.ac.za}
    \affiliation{Cosmology and Gravity Group, Department of Mathematics and Applied Mathematics, University of Cape Town, Rondebosch 7700, Cape Town, South Africa}

\author{ Peter Dunsby}
    \affiliation{Cosmology and Gravity Group, Department of Mathematics and Applied Mathematics, University of Cape Town, Rondebosch 7700, Cape Town, South Africa}
\affiliation{South African Astronomical Observatory, Observatory 7925, Cape Town, South Africa}
 \affiliation{Center for Space Research, North-West University, Potchefstroom 2520, South Africa}
\author{\'{A}lvaro de la Cruz-Dombriz}
      \affiliation{ Departamento de F\'{i}sica Fundamental, Universidad de Salamanca, 37008 Salamanca, Spain}
    \affiliation{Cosmology and Gravity Group, Department of Mathematics and Applied Mathematics, University of Cape Town, Rondebosch 7700, Cape Town, South Africa}

\date{\today}

\begin{abstract}
We explore the dynamics and gravitational-wave emission from black hole pairs on unbound orbits undergoing close hyperbolic encounters (CHEs) in dense astrophysical environments. While General Relativity predicts gravitational Bremsstrahlung radiation occurring at periastron, the contribution of the scalar gravitational-wave mode in fully general $f(R)$ gravity remains largely unexplored. We characterize the $f(R)$ scalar-mode gravitational radiation for both non-precessing and precessing hyperbolic orbits, identifying potential detection signatures for advanced gravitational-wave observatories. By systematically varying orbital precession and eccentricity, we examine their influence on the emitted gravitational waves. We derive potentially detectable time delay and scalar-to-tensor amplitude ratio estimates for representative astrophysical environments and determine optimal orbital configurations for the detection of $f(R)$ scalar gravitational waves from hyperbolic encounters. Our results provide the theoretical framework for scalar-mode signals and observables, establishing CHEs as a promising probe of $f(R)$ gravity with future detectors. 
\end{abstract}

\maketitle

\section{Introduction}
\label{sec:intro}
Since the first detection of gravitational waves in 2015, approximately 100 further detections of mer\-ging binary systems have been observed \cite{Broekgaarden_Banagiri_Payne_2024}. To date, these coalescing binaries have been the only astrophysical sources detected by gravitational-wave interferometers \cite{DePergola_Stuver_2024}.
The increasing sensitivity of Advanced LIGO-Virgo-KAGRA is likely to increase the rate of detection of binaries and broaden the scope of possible sources \cite{Abbott_Abbott_Abbott_Abraham_Acernese_Ackley_Adams_Adya_Affeldt_Agathos_etal._2020, PhysRevD.108.024013, PhysRevD.69.102001, PhysRevD.109.103004}. Complementarily, there is strong support for burst gravitational waves as the next target candidate for future gravitational-wave interferometer detections \cite{Powell_Lasky_2024, Garc_a_Bellido_2017, O’Leary_Kocsis_Loeb_2009, PhysRevD.109.042009, García-Bellido_Nesseris_2018, PhysRevLett.133.201401, PhysRevD.108.024013, Abadie_Abbott_Abbott_Abernathy_Accadia_Acernese_Adams_Adhikari_Affeldt_Allen_etal._2011, Dandapat_2023, Mukherjee_Mitra_Chatterjee_2021}.\\

Burst gravitational waves are attributed to many sources, including core-collapse supernovae \cite{PhysRevD.110.042007, Dimmelmeier_Ott_Marek_Janka_2008}, fast radio bursts, 
magnetars \cite{Christensen_2018, Abadie_Abbott_Abbott_Abernathy_Accadia_Acernese_Adams_Adhikari_Affeldt_Allen_etal._2011}, cosmic strings \cite{PhysRevLett.85.3761, PhysRevD.64.064008, PhysRevLett.125.211302} and scattering events caused by compact objects in dense environments 
interacting on unbound orbits \cite{Powell_Lasky_2024, PhysRevD.109.064001, Knee_McIver_Naoz_Romero-Shaw_Hoang_Grishin_2024, Xuan_Naoz_Kocsis_Michaely_2024}. In particular, some percentage of black-hole po\-pu\-la\-tions inhabiting globular clusters or AGN disks are expected to interact through orbits that result in capture with a highly eccentric orbit before circularization \cite{Lower_2018, BHsinGlobularClusters, Thorne_1995, Cheeseboro_Baker_2021}, as well as through so-called close hyperbolic encounters (CHEs). In these encounters, while interacting for a short time before continuing on 
their respective trajectories, black holes approach each other on unbound hyperbolic trajectories from large distances with tangential velocities large enough to prevent capture \cite{Capozziello_DeLaurentis_DePaolis_Ingrosso_Nucita_2008}. CHEs are also likely to occur within stellar po\-pu\-la\-tions orbiting black holes in central galactic regions and proposed primordial black-hole populations \cite{Powell_Lasky_2024, Garc_a_Bellido_2017}. Thus, CHEs can cause spin-induction \cite{Jaraba_García-Bellido_2021, Rodríguez-Monteverde_Jaraba_García-Bellido_2024}, subsequent mergers, contributions to the stochastic gravitational-wave background \cite{Caldarola_2024, Christensen_2018} and constitute a possible method for investigating dynamical friction due to dark matter \cite{Chowdhuri_2024}.\\

These interactions cause gravitational scattering, accelerating the masses and releasing burst gravitational waves in the form of Bremsstrahlung radiation, which generates waveforms with a characteristic peak at closest approach  \cite{Turner_1977, Turner_Will_1978}. The amplitude of these waveforms is highly dependent on the orbital parameters assumed, implying that detection of a CHE event would significantly improve our understanding of the properties and evolution of astrophysical environments containing populations of compact objects \cite{Caldarola_2024}. Additionally, studying these scattering events can provide valuable insights into the dynamics and properties of central galactic regions and AGN disks \cite{Amaro-Seoane_2018, Zhang_2020}.\\

Thus far, the literature has focused on CHEs in the context of General Relativity (GR); however, 
since gravitational waves in GR comprise only two tensor radiation modes, while many theories beyond the Eins\-teinian paradigm imply the existence of additional radiation modes,
the observation of gravitational waves from CHEs could provide tests of modified theories of gravity \cite{PhysRevD.77.024033, Yamada_Narikawa_Tanaka_2019}. In particular, $f(R)$ theories introduce one longitudinal scalar polarization mode for gravitational waves in addition to the usual transverse tensor modes already present in GR \cite{AparicioResco:2016xcm,
Ananda:2007xh,
Capozziello:2008fn,
Nishizawa:2014zra,
Bouhmadi-Lopez:2012piq,
Faraoni:1998sp,
Capozziello_2008, Liang_Gong_Hou_Liu_2017, Näf_Jetzer_2011, fan2024polarizationmodesgravitationalwaves}. While recent observations of gravitational waves have constrained gra\-vi\-ta\-tional-wave speed \cite{Ezquiaga_Zumalacárregui_2017, Creminelli_Vernizzi_2018}, the scalar propagation arising from several scalar-tensor theories, along with observable quantities predicted by these models, has remained mostly unconstrained. Additionally, cosmologically viable $f(R)$ theories that admit observationally compatible weak-field limits are those that concurrently exhibit $\Lambda$CDM expansion histories \cite{PhysRevD.91.103528} ({\it c.f.} \cite{Felice_Tsujikawa_2010, Sotiriou_Faraoni_2010} for comprehensive reviews on $f(R)$\,).\\

Since $f(R)$ models provide a compelling alternative explanation to late-time cosmic acceleration without the need of either {\it ad-hoc} dark energy or a cosmological constant \cite{Starobinsky_2007, PhysRevD.74.086005, lobo2008darkgravitymodifiedtheories}, investigating scattering events within the $f(R)$ framework may be used as a testbed for modified gravity theories while pushing the boundaries of our understanding of gravitational interactions in unbound orbits  \cite{Vainio_Vilja_2017}.\\

In this work, we investigate gravitational radiation generated by two CHE scenarios  in the realm of fully general $f(R)$ metric theories in which novel scalar-mode gravitational wave signals are present. The first case consists of a Newtonian CHE, where the relativistic effects of orbital precession are not considered, while the second case takes precession into account through the inclusion of an orbital precession parameter.\\

This communication is organized as follows: Section \ref{sec:gweq} outlines the theoretical foundations and the quadrupole radiation formulae for the canonical GR tensor modes,  along with the scalar mode arising from the linearized $f(R)$ field equations. Section \ref{sec:ches} analyzes the waveforms generated in both non-precessing and precessing cases. Constraints on the mass of the scalar mode are discussed in Section \ref{sec: scalarModeMass}, along with a derivation of observational estimates for time delay between the arrival of the tensor and scalar modes. The ratio of scalar to tensor amplitudes is investigated in Section \ref{sec: STRatio}. To conclude, Section  \ref{Conclusions} contains a summary of our work and future prospects. Throughout this work international standard units are used unless stated otherwise and we adopt the metric convention $\{-, +, +, +\}$.

\section{Gravitational Waves in f(\textit{R}) Gravity}
\label{sec:gweq}
We begin with the total $f(R)$ metric action,
\be\label{action}
S = \frac{1}{2 \kappa}\int {\rm d}^4 x\, \sqrt{-g}\,\, \big [\,f(R) +\mathcal{L}_{matter} \big]
\;,
\ee
where $\kappa \equiv \dfrac{8\pi G}{c^{4}}$, $G$ is the gravitational constant and $c$ the speed of light, $f(R)$ is a general function of the Ricci scalar $R$ and $\mathcal{L}_{matter}$ is a general matter Lagrangian. Varying \eqref{action} with respect to the metric tensor yields the $f(R)$ field equations, 
\begin{equation}
    \label{eom}
\mathcal{G}_{\mu\nu} =  \kappa\, T_{\mu\nu} \;,    \end{equation}
    where
\begin{equation}     \mathcal{G}_{\mu\nu} \equiv f'(R)R_{\mu\nu} 
        - \dfrac{1}{2}g_{\mu\nu}f(R)+(g_{\mu\nu}\Box - \nabla_{\mu}\nabla_{\nu})f'(R)\;,
    \end{equation}
Prime denotes derivative with respect to $R$ and $T_{\mu\nu}$ is the energy-momentum tensor derived from the variation of $\mathcal{L}_{matter}$, in \eqref{action}.\\

The wave equations are found through the perturbation of the metric tensor in \eqref{eom} around a Minkowski metric\footnote{While the background considered herein is Schwarzschild, the perturbed Minkowski background can be chosen instead for regions far from the Schwarzschild radius.} , 
\begin{equation}
\label{metric_pert}
    g_{\mu\nu} = \eta_{\mu\nu} + h_{\mu\nu}\; \quad \text{where}\quad   \vert h_{\mu\nu} \vert \ll 1 \;.
\end{equation}
The first order perturbations of $f(R)$ and $f'(R)$ around the background curvature,  $\overline{R} = 0$, are given by 
\begin{eqnarray}
    f(R) = f\left(0\right)+f'(0)\,\delta R \;,\label{perturbationsofF}\\
    f'(R) = f'\left(0\right)+f''(0)\,\delta R\;\label{perturbationFF}\,.
\end{eqnarray}

Since \eqref{eom} contains in principle both the tensor and scalar modes, we define
\begin{equation}\label{PhiEqn}
   \Phi \equiv \dfrac{f''(0)}{f'(0)}\,\delta R\;,
\end{equation}

where $\Phi$ represents the scalar mode gravitational radiation and allows us to perform the separation of the tensor and scalar modes in the resulting wave equations. To this end, we perform the following transformation on the metric perturbation \cite{Berry_Gair_2012, PhysRevD.108.024003}
\begin{equation}\label{coordtransform}
\tilde{h}_{\mu\nu} = h_{\mu\nu} + \left(\beta\,\Phi - \dfrac{1}{2}h\right)\eta_{\mu\nu}\,, 
\end{equation}
which includes the trace of the metric perturbation, $h$, and the dimensionless parameter $\beta$, a constant whose chosen value will later allow us to isolate the scalar and tensor modes from \eqref{coordtransform}.\\ 

We now perform a gauge transformation by employing the Lorentz gauge,
\begin{equation}
    \partial^\mu \tilde{h}_{\mu\nu} = 0 \;.
\end{equation}

Subject to this choice, the Ricci tensor and Ricci scalar to lowest order are given by 
\cite{Casado-Turrion:2024esi}
\begin{align}
\label{Rtensor}
    \delta R_{\mu\nu} = -\dfrac{1}{2}\Box\left(\tilde{h}_{\mu\nu} - \dfrac{1}{2} \eta_{\mu\nu}\tilde{h} \right)+\beta\left(\eta_{\mu\nu}\Box\Phi+2\partial_{\mu}\partial_{\nu} \Phi \right)\,,\nonumber\\
\end{align}
and 
\begin{equation}\label{R}
    \delta R = \dfrac{1}{2}\Box \tilde{h} - 3\beta\Box \Phi\;.
\end{equation}
Since we are performing our calculations in an empty (vacuum) background, we set $\overline{T}_{\mu\nu} = 0$ and thus substi\-tu\-ting \eqref{Rtensor} and \eqref{R} into \eqref{eom} yields the perturbed vacuum field equations, 
\begin{equation}\label{Linearizedf(R)}
    -\dfrac{1}{2}\Box \tilde{h}_{\mu\nu} +(\beta+1)\left[\eta_{\mu\nu}\Box - \partial_{\mu}\partial_{\nu}\right]\Phi = 0\;.
\end{equation}

Now, setting $\beta = - 1$ removes the scalar mode contribution to \eqref{Linearizedf(R)}, yielding the tensor mode wave equation,
\begin{equation}\label{TensorWaveEquation}
    \Box \tilde{h}_{\mu\nu} = 0\;,
\end{equation}
which is fully analogous to the  standard GR prediction. 
Extracting the scalar mode wave equation, however, requires further calculations. We begin by taking the trace of the vacuum $f(R)$ field equations \eqref{eom}, which gives
\begin{equation}
\label{f(R)Trace} 
    \Box f'(R) = \dfrac{1}{3}\left( 2f(R) - Rf'(R) \right)\;.    
\end{equation}

Then performing the metric perturbation
\eqref{metric_pert} on \eqref{f(R)Trace} and substituting \eqref{perturbationsofF}, \eqref{perturbationFF} and \eqref{PhiEqn}, the scalar mode wave equation becomes
\begin{equation}\label{ScalarWaveEquation}
    \left(\Box - m_{\text{S}}^{2}\right)\Phi = 0 \;,
\end{equation}
where 
\begin{equation}\label{ms_eqn}
    m_{\text{S}}^{2} = \dfrac{1}{3} \dfrac{f'(0)}{f''(0)}\;,
\end{equation}
is usually under\-stood as the squared mass of the pro\-pa\-gating scalar field \footnote{provided that $f''(0)\neq 0$ to avoid pathological scenarios.} \cite{Casado-Turrion:2023rni}, and determines the range of the scalar force \cite{Felice_Tsujikawa_2010}. Additionally, from \cite{Faraoni_2004}, we see that $m_{\text{S}}$ is canonically expressed in units of $L^{-2}$.

\subsection{Solutions of the field equations}
\label{Sec:II:A}
We now focus on solutions to the respective wave equations, beginning with the tensor modes.
\subsubsection{Solution of the tensor mode wave equation}
We begin by introducing the energy-momentum tensor as a source to \eqref{TensorWaveEquation} \cite{Maggiore_2007},  
\begin{equation}\label{tensorwaveeq}
    \Box \tilde{h}_{\mu\nu} = \frac{8\pi \textbf{G}}{c^4} \,T_{\mu\nu}\;, 
\end{equation}
where $\textbf{G}$ is the effective $f(R)$ gravitational constant defined as $G/f'(0)$, and introduce a Green's function as the solution to this sourced wave equation, 
\begin{equation}\label{tensorgreensfunction}
    \Box \mathcal{G}_{\rm T}(\textbf{x}) = \delta(\textbf{x})\;.
\end{equation}
The Fourier representation of $\tilde{h}_{\mu\nu}$ is given by
\begin{equation}\label{tensorFourier}
   \tilde{h}_{\mu\nu} (\textbf{x}, t) = \frac{1}{\sqrt{2\pi}}\int {\rm d}k^0\, \tilde{h}_{\mu\nu}(\textbf{x}, k^0)\,{\rm e}^{-ik^0 t}, 
\end{equation}

where $k^0 \equiv \dfrac{\omega}{c}$. Now, using the Fourier representation of the Green's function we find the solution to \eqref{tensorwaveeq} as
\begin{align}\label{tensorwaveeqsolution}
    \tilde{h}_{\mu\nu} (\textbf{x}, k^0) = -\frac{16\pi \textbf{G}}{c^4} \int {\rm d}^3 \textbf{x}\, \mathcal{G}_{\rm T} (\Delta\textbf{x}, k_0)\,T_{\mu\nu}(\textbf{x}', k_0)\;,
\end{align}
where the Fourier representation of the Green's function is
\begin{align}
    \mathcal{G}_{\rm T}(\textbf{x}, k_0) &= \frac{1}{\sqrt{\pi}^3} \int \frac{1}{\sqrt{\pi}^3} \frac{1}{-\textbf{k}^2 +k_0^2}{\rm e}^{i\textbf{k}\cdot\textbf{x}}\,{\rm d}^3\textbf{k}\; \nonumber\\
    &= - \frac{1}{4\pi |\textbf{x}|}\, {\rm e}^{ik_0 |\textbf{x}|}\;.\label{tensorgreensFourier2}    
\end{align}
After substituting \eqref{tensorgreensFourier2} into \eqref{tensorwaveeqsolution} and finally into \eqref{tensorFourier}, we obtain the time-retarded solution to \eqref{tensorwaveeq}:
\begin{equation}\label{TensorWaveSolution}
\tilde{h}_{\mu\nu} (\textbf{x}, t) =\frac{4 \textbf{G}}{c^4}\int {\rm d}^3\textbf{x}'\, \frac{T_{\mu\nu}\big(\textbf{x}', t-|\frac{\Delta\textbf{x}}{c}| \big)}{|\Delta\textbf{x}|} \;,
\end{equation}
where $\Delta\textbf{x} \equiv \textbf{x} - \textbf{x}'$. We see from this solution that,  in agreement with current observations \cite{Creminelli:2017sry,Ezquiaga:2017ekz}, the tensor mode gravitational radiation travels at the speed of light. Note also that the $f(R)$ contribution to this solution depends on the effective gravitational constant $\textbf{G}$, and GR is recovered in the limit where $\textbf{G}$ is set to $G$, i.e., $f(R)=R$.\\

\subsubsection{Solution of the scalar mode wave equation}

Similarly, \eqref{ScalarWaveEquation} is sourced by the trace of the energy-momentum tensor, 
\be\label{sourcedscalarwaveeq}
   (\Box - m_{\text{S}}^2)\Phi =  \frac{8\pi\textbf{G}}{c^4}\,T\;.
\ee
To solve this wave equation we use the retarded Green's function, $\mathcal{G}_S$, satisfying 
\begin{equation}\label{ScalarGreensFunc}
    (\Box - m_{\text{S}}^2)\,\mathcal{G}_{\rm S}{(\textbf{x})} = \delta(\textbf{x})\;.
\end{equation}
Note that the inclusion of the mass $m_\text{S}$ in \eqref{ScalarGreensFunc} differentiates this Green's function from that used for the tensor mode solution \cite{PhysRevD.108.024003}.\\

Integrating \eqref{ScalarGreensFunc} with respect to the wave vector yields the Fourier representation of the Green's function \eqref{sourcedscalarwaveeq} \cite{Arfken_Weber_Harris_2013, PhysRevD.108.024003} and is given by
\begin{align}
\mathcal{G}_{\rm S}(\textbf{x}, k_0) &= \frac{1}{\sqrt{2\pi}^3} \int\frac{{\rm e}^{i\textbf{k}\cdot\textbf{x}}}{-\textbf{k}^2 - m_{\text{S}}^2 + k_0^2}\,{\rm d}^3\textbf{k}\nonumber\\
    &= -\frac{1}{4\pi|\textbf{x}|}\,{\rm e}^{i\sqrt{k_0^2 - m_{\text{S}}^2}|\textbf{x}|}\;.
\end{align}
where $k_0 \equiv \dfrac{\omega}{c}$. The Fourier representation of the scalar mode is thus given by
\begin{equation}
\Phi(\textbf{x}, k^0) = -\frac{16\pi\textbf{G}}{c^4} \int {\rm d}^3\textbf{x}'\,\mathcal{G}_{\rm S}(\textbf{x}-\textbf{x}', k_0)\,T(\textbf{x}', k_0)\;,
\end{equation}
and consequently, 
the inverse Fourier transformation of the scalar mode becomes
\begin{align}
    \Phi(\textbf{x}, t)&= \frac{4\textbf{G}}{c^4}\int {\rm d}^4 \textbf{x}'\, T(\textbf{x}', t')\,\mathcal{G}_{\rm S}(|\textbf{x}-\textbf{x}'|, \Delta t)\;.\label{phigreen} 
\end{align}
where $\Delta t \equiv t-t'$. Whence, the Green's function Fourier representation yields
\begin{multline}\label{greensfuncFourier}
    \mathcal{G}_{\rm S}(|\Delta\textbf{x}|, \Delta t) = \frac{1}{2\pi |\Delta\textbf{x}|}\\ 
    \times\int_{-\infty}^{\infty} {\rm e}^{i\left(\sqrt{k_0^{2}-m_{\text{S}}^2}|\frac{\Delta\textbf{x}}{c}|-k_0 \Delta t\right) }{\rm d}k^0\;.
\end{multline}
Following the methodology presented in \cite{PhysRevD.108.024003, Näf_Jetzer_2011}, Eq. \eqref{phigreen} can be represented using a Bessel Function as
\begin{multline}\label{besselfunc} 
\mathcal{G}_{\rm S}(|\Delta\textbf{x}|, \Delta t) = \frac{\delta(\Delta t - |\Delta\textbf{x}|)}{|\Delta\textbf{x}|}\\ 
        - \frac{m_{\text{S}} \,J_1 \left( m_{\text{S}} \sqrt{\Delta t^2 - |\frac{\Delta\textbf{x}}{c}|^{2}}\right)}{\sqrt{\Delta t^2 - |\frac{\Delta\textbf{x}}{c}|^{2}}}\theta(\Delta t - \Big|\frac{\Delta\textbf{x}}{c}\Big|)\;.
\end{multline}
Now substituting \eqref{besselfunc} into \eqref{phigreen}, the retarded time solution of the scalar mode is found to be
\begin{align}\label{retardedsolution}
    &\Phi (\textbf{x}, t)\,=\, \frac{4\textbf{G}}{c^4} \int {\rm d}^3\textbf{x}' \Bigg [ \frac{T(\textbf{x}', t-|\frac{\Delta\textbf{x}}{c}|)}{|\Delta\textbf{x}|}\nonumber\\ 
    &-\int_{-\infty}^{t-|\frac{\Delta\textbf{x}}{c}|} {\rm d}t' \frac{m_{\text{S}} J_1 \big( m_{\text{S}} \sqrt{\Delta t^2 - |\frac{\Delta\textbf{x}}{c}|^{2}}\big)}{\sqrt{\Delta t^2 - |\frac{\Delta\textbf{x}}{c}|^{2}}} T(\textbf{x}', t') \Bigg ]\;.
\end{align}

Defining the following simplifying parameters, 
\begin{align}
    t_p &= t-\Big|\frac{\Delta\textbf{x}}{c}\Big|\;,\\
    t_f &= t+\Big|\frac{\Delta\textbf{x}}{c}\Big|\;,\\
    \tau &= \sqrt{(t'-t_p)(t'-t_f)}\;,
\end{align}
where $t_p$ and $t_f$ represent the present and future times respectively, allows \eqref{retardedsolution} to be rewritten as 
\begin{align}
\Phi (\textbf{x}, t)\,&=\, 
    \frac{4\textbf{G}}{c^4}\int_V {\rm d}^3\textbf{x}' \nonumber\\
    & \times \int_0^{\infty} {\rm d}\tau \frac{J_1(m_{\text{S}}\,\tau)\,T\left(\textbf{x}', t-\sqrt{\tau^2+|\frac{\Delta\textbf{x}}{c}|^{2}}\right)}{\sqrt{\tau^2 + |\frac{\Delta\textbf{x}}{c}|^{2}}}.\nonumber\\
\end{align}
Finally, introducing the following transformation for the integration variable $\tau$ 
\begin{equation}
    \tau = \Big|\frac{\Delta\textbf{x}}{c}\Big| \sinh{\xi}\;,
\end{equation}
produces the solution to the scalar mode wave equation, given by
\begin{align}
\label{ScalarWaveSolution}
\Phi(\textbf{x}, t) = \frac{4\textbf{G}}{c^4}\int {\rm d}^3\textbf{x}' &\int_0^{\infty}{\rm d}\xi \left[ \frac{\delta (\xi)}{|\Delta\textbf{x}|} - J_{1}(m_{\text{S}} |\Delta\textbf{x}|\sinh{\xi})\right]\nonumber\\ 
&\times T(\textbf{x}', t-\cosh{\xi \Big|\frac{\Delta\textbf{x}}{c}\Big|})\;.
\end{align}
From \eqref{ScalarWaveSolution}, we note that in the integration interval, $\cosh{\xi}$ takes values in the range from 1 to $\infty$ and the velocity of the scalar mode propagation is given by $c_{\text{S}} \equiv 1/\cosh\xi$.
When $\xi = 0$, the scalar mode pro\-pa\-gates at the speed of light ($c=1$ here), and in the limit $\xi \rightarrow \infty$ the scalar mode does not pro\-pa\-gate, i.e., $c_s \rightarrow 0$ \cite{PhysRevD.108.024003}.

\subsection{Quadrupole radiation formulae}
\label{Sec:II:B}
We now focus on the quadrupole radiation formulae for both the scalar and tensor modes, beginning with the tensor modes.

\subsubsection{Tensor mode quadrupole radiation}
We begin by integrating \eqref{TensorWaveSolution}, assuming that the source and observer are significantly distant and that the source is non-relativistic. This assumption leads to the simplification
\begin{equation}
\label{curly_R}
|\Delta\textbf{x}| \sim |\textbf{x}| \equiv \mathcal{R} \;, 
\end{equation}
Thus, to leading order, \eqref{TensorWaveSolution} can be expressed as
\begin{equation}\label{Barhij}
    \tilde{h}^{ij}(\textbf{x}, t) \sim \frac{4\textbf{G}}{\mathcal{R}c^4} \int T^{ij} (\textbf{x}', t-\mathcal{R})\,{\rm d}^3\textbf{x}' \;.
\end{equation}

Using conservation of the energy-momentum tensor $\big( \nabla_{\mu} T^{\mu\nu} = 0 \big)$ yields the relation 
 
\begin{equation}
    \partial_\mu\partial_\nu x^i x^j T^{\mu\nu}(x) = 2T^{ij}(x) \;,
\end{equation}
which, after spatially integrating and neglecting surface terms, yields
\begin{equation}\label{2Tij}
    2 \int T^{ij}(x)\,{\rm d}^3 \textbf{x} = \partial_0 \partial_0 \int {\rm d}^3 \textbf{x} \hspace{0.1cm}x^i x^j T^{00}(x)\;. 
\end{equation}
Substituting \eqref{2Tij} into \eqref{Barhij} and simplifying gives the tensor mode solution expressed in terms of the second time derivative of the quadrupole moment, $Q^{ij}$, as
\begin{align}\label{TensorModeQuadrupoleRadiationFormula}
    \tilde{h}^{ij}(\textbf{x}, t) &= \frac{2 \textbf{G}}{\mathcal{R}c^4} \ddot{Q}^{ij} \;.
\end{align}
We note here that the projection operators are necessary to describe the polarization of the tensor modes \cite{Smarr_1979}.
Consequently, the decomposition of \eqref{TensorModeQuadrupoleRadiationFormula} results in the radiation formulae for the two tensor polarization modes, $h_+$ and $h_{\times}$, as
\begin{align}
    \tilde{h}_{+} &= \frac{\mathbf{G}}{\mathcal{R}c^4} \left ( \ddot{Q}^{11} - \ddot{Q}^{22}\right ) \;,\label{h_plus_fR}\\
    \tilde{h}_{\times} &= \frac{2 \mathbf{G}}{\mathcal{R}c^4}\ddot{Q}^{12}  \;\label{h_cross_fR}.
\end{align}

\subsubsection{Scalar mode quadrupole radiation}
For the scalar mode formula, we begin with a variable change that simplifies the integration of the Bessel function in \eqref{ScalarWaveSolution} \cite{PhysRevD.108.024003}, 
\begin{equation}
    w \equiv m_{\text{S}} \mathcal{R}\sinh{\xi}\;.
\end{equation}
It follows then that the second integration term becomes
\begin{align}\label{variablechangedIntegral}    \int_{0}^{\infty} {\rm d}\xi\, m_{\text{S}} J_1(m_{\text{S}}\mathcal{R}\sinh{\xi}) &= \int_{0}^{\infty} {\rm d}w\, \frac{m_{\text{S}} J_1(w)}{\sqrt{w^2 + (m_{\text{S}} \mathcal{R})^2}}\;,
\end{align}
with
\begin{align}
\int_{0}^{\infty} {\rm d}w\, \frac{m_{\text{S}} J_1(w)}{\sqrt{w^2 + (m_{\text{S}}\mathcal{R})^2}} &= m_{\text{S}}\, I_{\frac{1}{2}}\left(\frac{m_{\text{S}} \mathcal{R}}{2}\right) K_{\frac{1}{2}}\left(\frac{m_{\text{S}} \mathcal{R}}{2}\right)\;,
\end{align}
where $I_{\frac{1}{2}}$ and $K_{\frac{1}{2}}$ are modified Bessel functions satisfying \cite{PhysRevD.108.024003}
\begin{equation*}
I_{\frac{1}{2}}(z) = \sqrt{\frac{2}{\pi z}}\sinh{z}
   \quad\text{and}\quad 
K_{\frac{1}{2}}(z) = \sqrt{\frac{\pi}{2z}}\,{\rm e}^{-z} \;.
\end{equation*}
Assuming that the scalar mode propagation speed is approximately constant for distant observers and that the energy-momentum tensor is not dependent on the value of $\cosh{\xi}$ (i.e., it is independent of the scalar mode group velocity), \eqref{ScalarWaveSolution} becomes
\begin{align}
    \Phi &= \frac{4\textbf{G}\,{\rm e}^{-m_{\text{S}}\mathcal{R}}}{\mathcal{R}c^4}\int {\rm d}^3 \textbf{x}' \,T\left( \textbf{x}', t-\frac{\mathcal{R}}{c_\text{S}}\right)\;,
\label{Phi_fR}
\end{align}
 
where $c_\text{S}$ is the propagation speed of the scalar mode.\\

In the above, the trace of the energy-momentum tensor $T$ can be split into spatial and temporal parts, $T = -T^0_{\;\;0} +T^i_{\;\;i}$. The integration of the spatial part, $T^i_{\;\;i}$, gives the second time derivative of the traced second mass moment, defined throughout the following as $\ddot{Q}$,  and the integration of the temporal part, $T^0_{\;\;0}$, gives the energy density defined throughout as $Mc^2$ where $M$ is the total mass of the system. Thus, the quadrupole radiation formula for the scalar mode is given by:
\begin{equation}
\label{scalar_Quadrupole_Formula}
    \Phi = \frac{4 \textbf{G}\,{\rm e}^{-m_\text{S}\mathcal{R}}}{\mathcal{R}c^2} M + \frac{2 \textbf{G}\,{\rm e}^{-m_\text{S}\mathcal{R}}}{\mathcal{R}c^4} \ddot{Q} \;,\\
\end{equation}

Thus, according to \eqref{scalar_Quadrupole_Formula},
the scalar mode exhibits a Yukawa-like potential behavior that depends on both $M$ and $\ddot{Q}$. As discussed in \cite{PhysRevD.108.024003}, this is a straighforward consequence of the non-vanishing mass of the scalar mode.


\section{Close hyperbolic encounters} \label{sec:ches}
\subsection{Hyperbolic Encounters in the Newtonian Limit of $f(R)$}
\label{sec:nonpre}
\begin{figure}
\centering
\includegraphics[width = 0.5\textwidth]{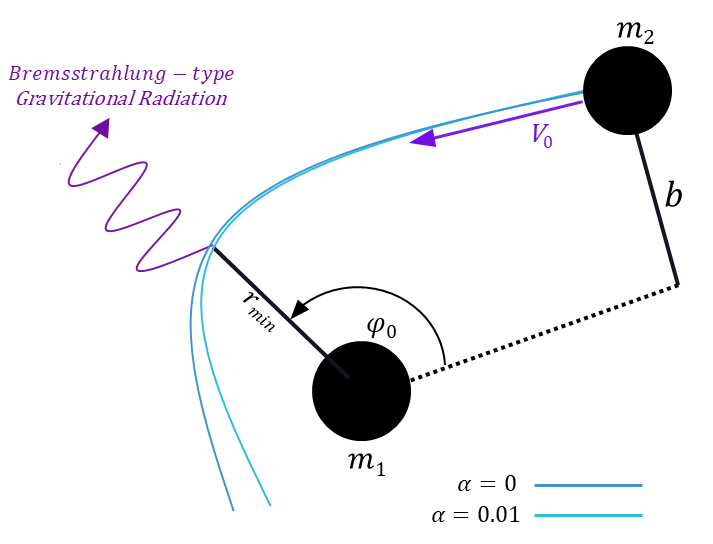}
\caption{Schematic representation of close hyperbolic encounter of masses $m_{1}$ and $m_{2}$, the latter with velocity $v_{0}$. The configuration is settled with an impact parameter $b$, periastron distance $r_{min}$ and periastron angle $\varphi_{0}$. The relativistic effect of precession is shown by the differing orbits identified by two values of the precession parameter $\alpha$, with the Newtonian case represented by $\alpha = 0$.}
\label{fig:KeplerCHEfig}
\end{figure}

We begin our analysis by considering the hyperbolic orbit of one black hole, $m_2$, approaching a companion black hole, $m_1$,
with asymptotic velocity $v_0$, impact parameter $b$, periastron distance $r_{min}$ and periastron angle $\varphi_0$. A visual representation of this configuration is shown in Fig. \ref{fig:KeplerCHEfig}.
In polar coordinates this orbit can be described through a Newtonian trajectory given by
\begin{equation}
    r(\varphi) = \frac{a(e^2 -1)}{1+e \cos{\varphi}}\;,
\end{equation}
 with $a=\text{\bf G}M/v_0^2$ and eccentricity $e>1$ to ensure hyperbolic trajectory and related to other orbital parameters through 

 \begin{equation} \label{ecc}
     e = \sqrt{1+\frac{b^2}{a^2}} = \sqrt{1+ \frac{b^2 v_0^4}{\textbf{G}^2 M^2}}
     >1\;,
 \end{equation}

where the total mass $M$ is defined as $M = m_1+m_2$.
The perisastron distance is given by
\begin{eqnarray}
 r_{min} = a(e-1)\;.
\end{eqnarray}
Planar motion of the black holes implies the conservation of the angular momentum, yielding the useful relation
\begin{equation}
    \dot{\varphi} = \frac{b v_0}{r^2(\varphi)} =\frac{\sqrt{\textbf{G}Mr_{min}(1+e)}}{r^2(\varphi)}\;.
\end{equation}
In the following, we choose a coordinate system where the trajectory is described by 
\be
 \vec{r} = r(\varphi)\left( \cos{\varphi(t)}, \sin{\varphi(t)},  0\right)
\ee
corresponding to a choice of the xy-plane as the plane of interaction with an observer in the z-direction. Thus, the quadrupole moment and its trace become
\begin{equation}
    Q^{ij} = \mu\, r^2(\varphi)
    \begin{pmatrix}
        \cos^2{\varphi} & \cos{\varphi}\sin{\varphi} & 0 \\
        \cos{\varphi}\sin{\varphi} & \sin^2{\varphi} & 0 \\
        0 & 0 & 0\\
    \end{pmatrix}\;,
\end{equation}
and
\begin{equation}
    Q = \mu\, r^2(\varphi) \;,
\end{equation}

respectively, where $\mu$ is the reduced mass of the system given by $\mu = \dfrac{m_1m_2}{M}$ and $Q$ is trace of the second mass moment.\\

\begin{figure}[htb]
    \centering
\includegraphics[width = 0.95\linewidth, height = 0.7\linewidth]{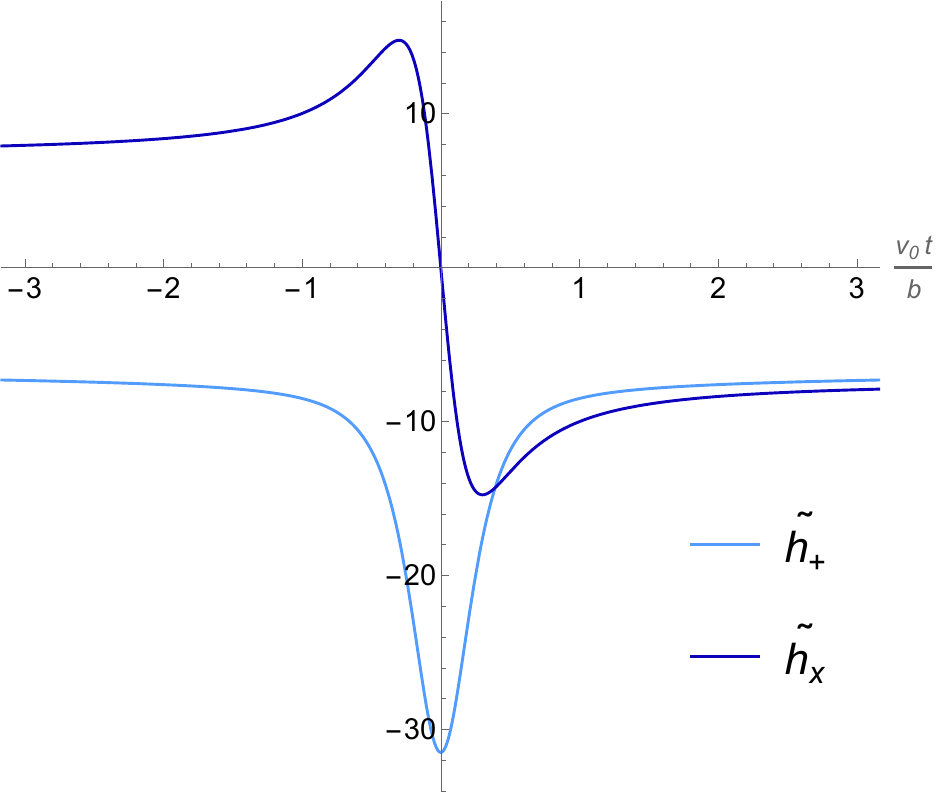}
\caption{Gravitational-wave strain of the tensor mode polarizations, $h_+$ and $h_{\times}$ according to 
 \eqref{eq: quadRad_hPlus} and 
\eqref{eq: quadRad_hCross} as functions of the dimensionless variable $\frac{v_0  t}{b}$. The strains are scaled by the constant prefactor $\frac{\text{\bf{G}}\mu v_0^{2}}{\mathcal{R}c^4}$.}
    \label{hplus_hcross_NO_prec}
\end{figure}

\begin{figure}[htb]
    \centering
    \includegraphics[width = 0.95\linewidth]{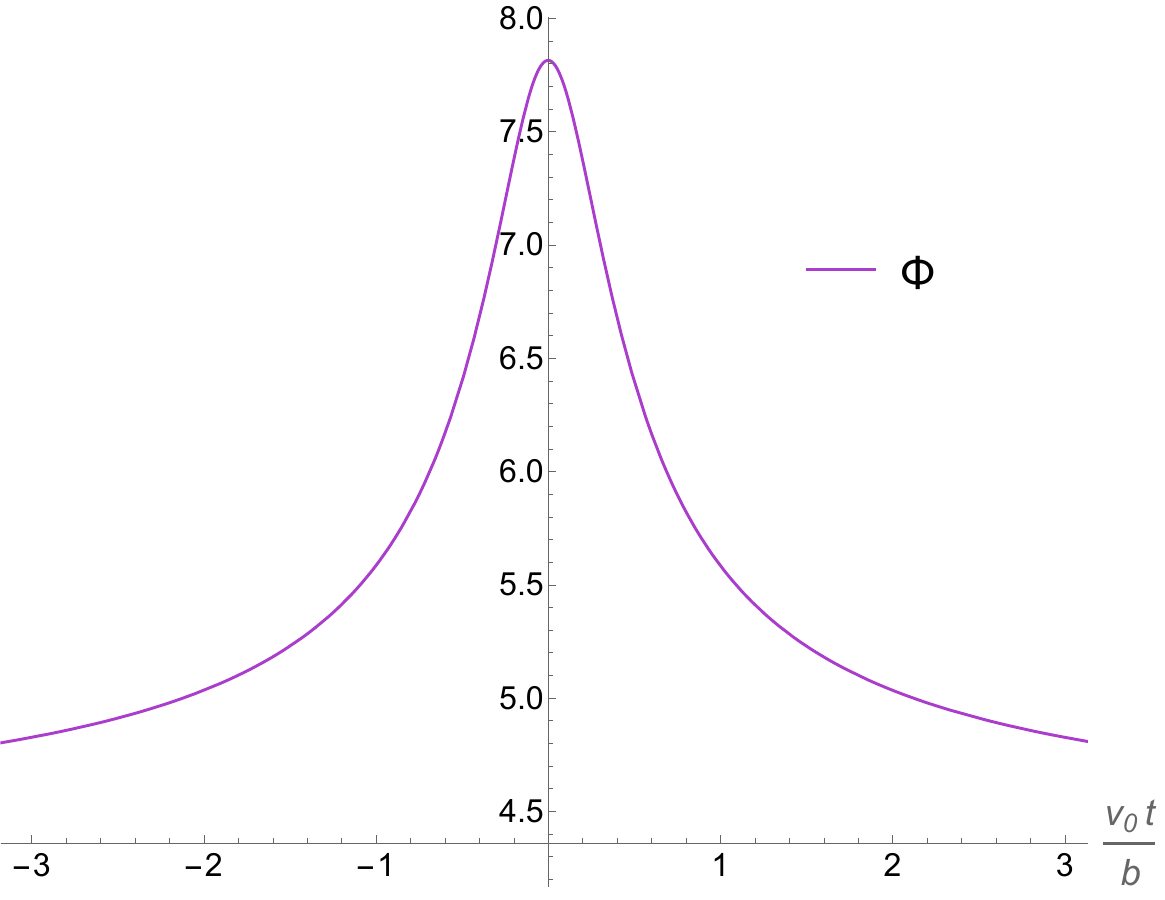}
\caption{Gravitational-wave strain of the scalar radiation mode $\Phi$ according to
\eqref{eq: quadRad_Phi} as a function of the dimensionless variable $\frac{v_0  t}{b}$. The strain is scaled by the constant prefactor $\frac{4\textbf{G}}{{\mathcal R} c^4}$, and $m_S = 1\times10^{-33} \,\,\, \text{eV}/c^2$.}   \label{fig:Phi_NO_prec}
\end{figure}

Substituting the above into \eqref{h_plus_fR}, \eqref{h_cross_fR} and \eqref{scalar_Quadrupole_Formula} yields the quadrupole radiation equations for the three polarization modes as
\begin{align}
    \label{eq: quadRad_hPlus}
    \tilde{h}_{+} \,=\,&\frac{2\textbf{G} v_0^2 \mu}{ (1-e^2)\mathcal{R}c^4}\,\bigg[ 5e\cos{\varphi} + 4\cos{2 \varphi} \notag \\
    &+e(2e + \cos{3 \varphi})\bigg]\;, \\
  \label{eq: quadRad_hCross}  %
    \tilde{h}_{\times} \,=\,& \frac{2\textbf{G} v_0^2 \mu}{ (1-e^2)\mathcal{R}c^4} \, \bigg[4\cos{\varphi} + e(3 + \cos{2\varphi})\bigg]\sin{\varphi} \:,
\\
\label{eq: quadRad_Phi}
    \Phi \,=\, & \frac{4  \textbf{G}\,{\rm e}^{-m_\text{S} \mathcal{R}}}{\mathcal{R}c^2} \left[ M +\frac{e\mu v_0^2 (e+\cos{\varphi})}{c^2(e^2-1)} \right]\;.
\end{align}
Figs. \ref{hplus_hcross_NO_prec} and \ref{fig:Phi_NO_prec} show the strain of the tensor and scalar modes, according to \eqref{eq: quadRad_hPlus}-\eqref{eq: quadRad_hCross} and \eqref{eq: quadRad_Phi} respectively, with respect to the dimensionless variable $v_0 t/b$ \cite{García-Bellido_Nesseris_2018}.
The amplitudes therein are plotted using parameter values characteristic for astrophysically-viable CHE scenarios, with the masses of each black hole set to 30$M_\odot$, the tangential velocity of the incoming mass $v_0 =0.007c$ and the impact parameter $b = 3.39\times10^{-14}\,\text{Mpc}$ respectively, and the orbital eccentricity chosen to be $e =  1.3$.
The distance from the observer to the system is set to 30 ${\rm Mp c}$ and $c$ has 
recovered its usual value in SI units. 
In addition, the plotted amplitudes have been scaled by their respective constant pre-factors.

We notice that for all modes, Bremsstrahlung-type radiation occurs at periastron. The scalar and the plus polarization modes have a cons\-tant non-zero strain far from the periastron distance, with strain increasing closer to periastron and peaking at periastron ($r_{min}$) - corresponding to $v_0 t / b =0$ as seen in Figs. \ref{hplus_hcross_NO_prec} and \ref{fig:Phi_NO_prec}, while the tensor cross polarization mode has a constant non-zero strain far from $r_{min}$, and zero strain at $r_{min}$ - consistent with findings in \cite{Caldarola_2024}. 


\subsection{Effects of varying the orbital eccentricity}\label{sec:ecc}
\begin{figure}
    \centering
    \includegraphics[width = 1.07\linewidth]{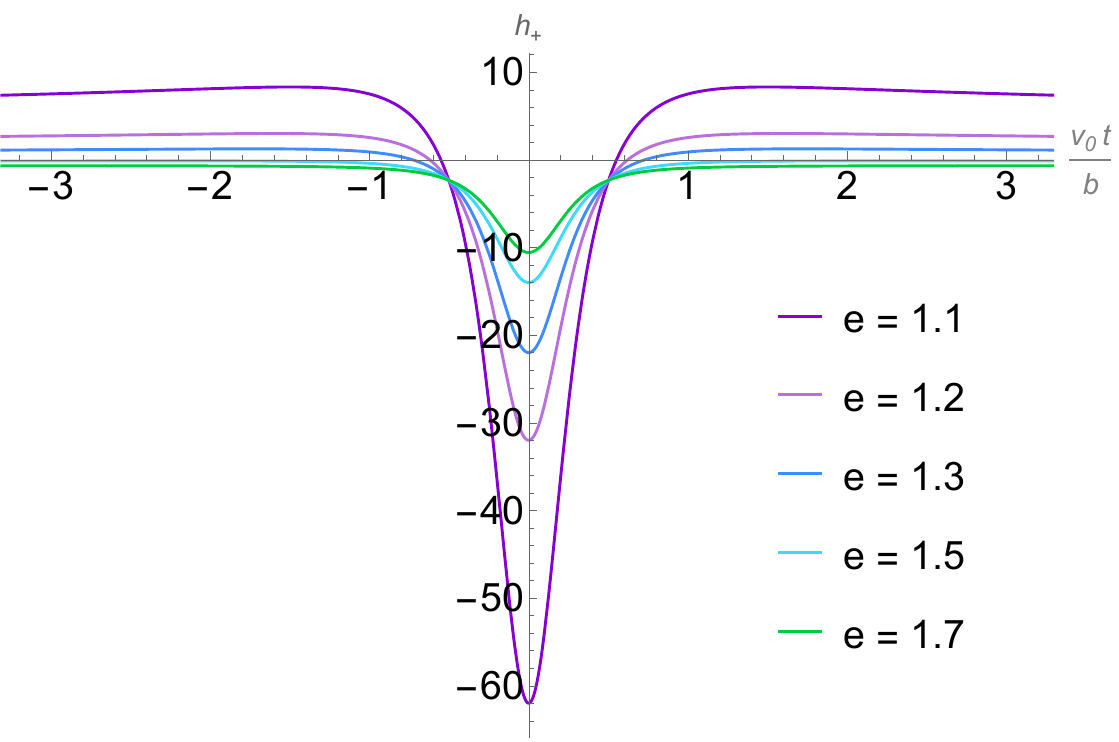}
    \caption{$h_{+}$ strain with varying eccentricities as a function of the dimensionless variable $\frac{v_0 t}{b}$. Strains are scaled by the constant prefactor $\frac{\text{\bf G}\mu v_0^{2}}{\mathcal{R}c^4}$.}
    \label{fig: hplus_E}
\end{figure}

\begin{figure}
    \centering
    \includegraphics[width = 1.07\linewidth]{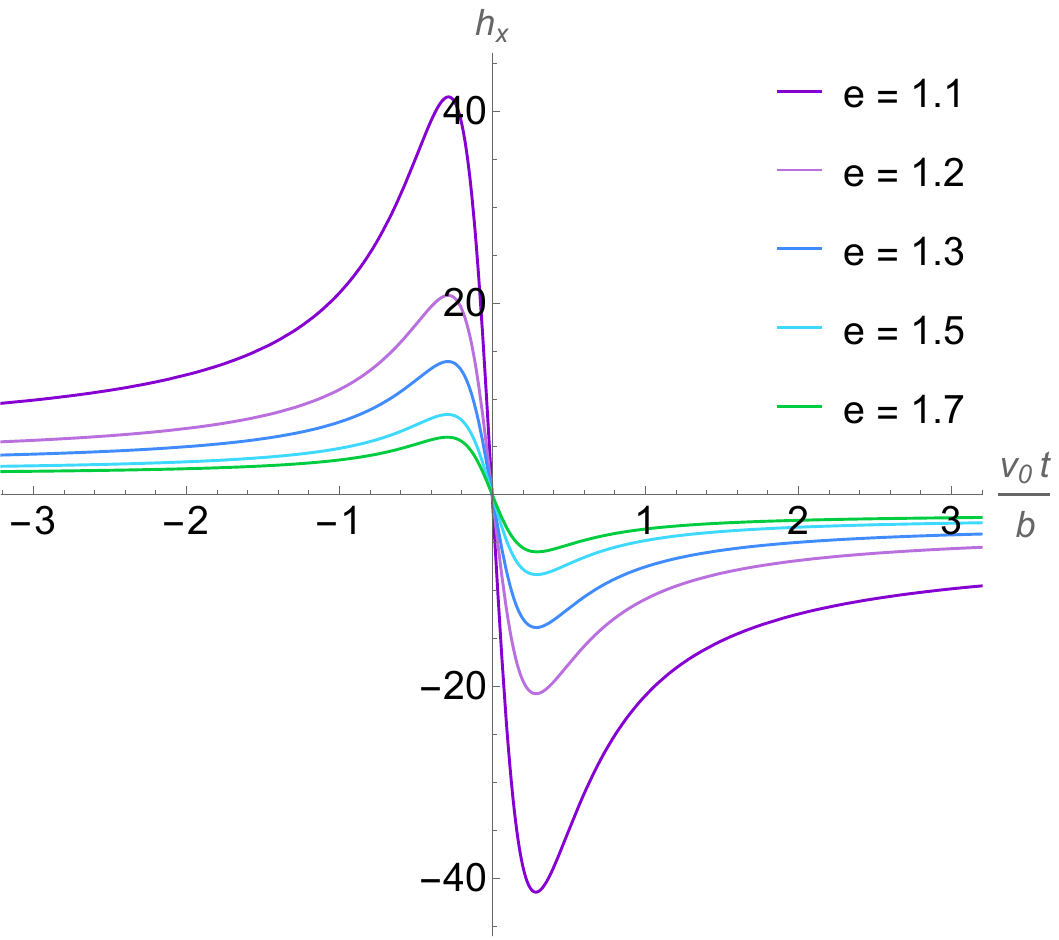}    \caption{$h_{\times}$ strain with varying eccentricities as a function of the dimensionless variable $\frac{v_0 t}{b}$. Strains are scaled by the constant prefactor $\frac{\text{\bf G}\mu v_0^{2}}{\mathcal{R}}$.}
    \label{fig:hcross_E}
\end{figure}

\begin{figure}
    \centering
    \includegraphics[width = 1.07\linewidth]{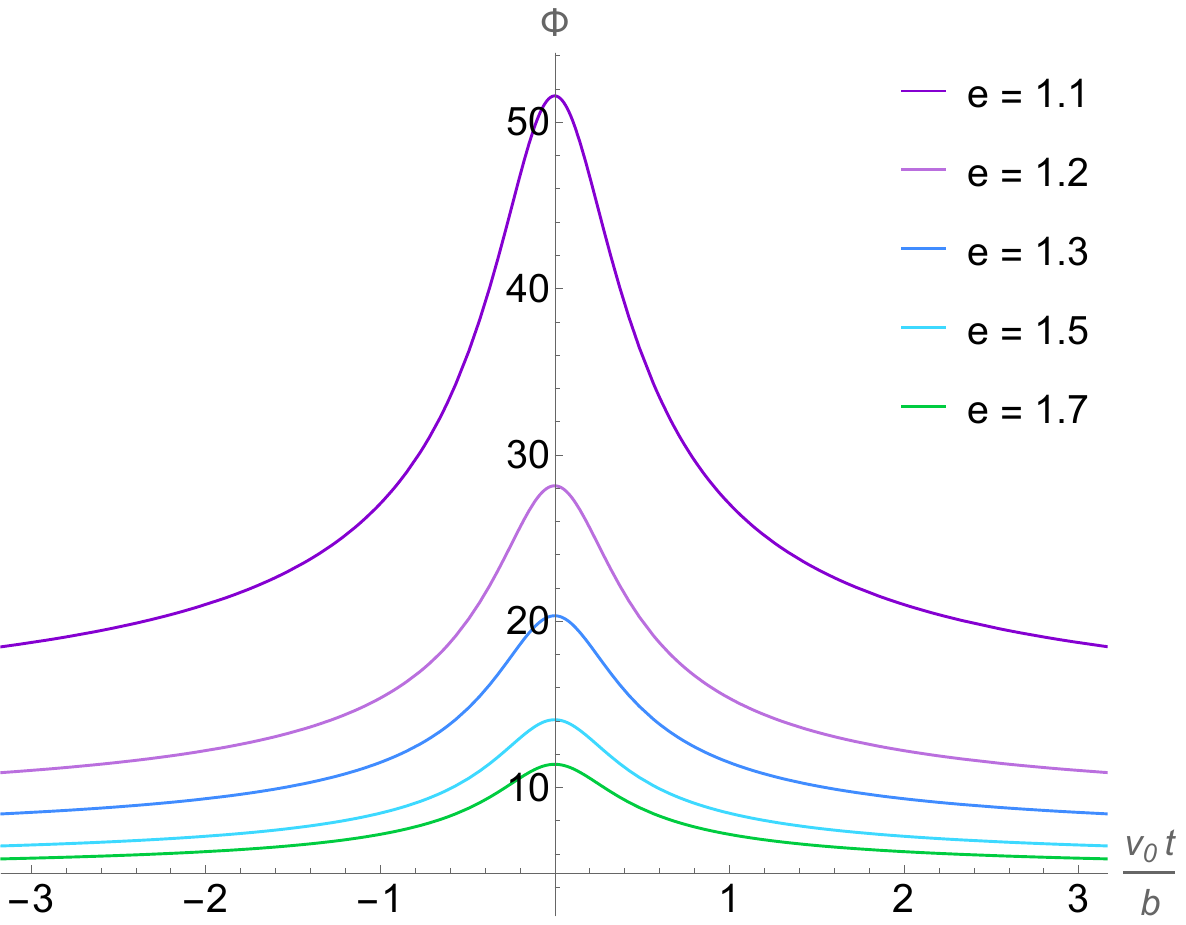}
    \caption{Scalar strain with varying eccentricities as a function of the dimensionless variable $\frac{v_0 t}{b}$. Strains are scaled by the constant prefactor $\frac{4\textbf{G}}{Rc^4}$, and $m_S = 1\times10^{-33} \,\,\, \text{eV}/c^2$.}
    \label{fig:Phi_E}
\end{figure}

Revisiting the quadrupole radiation expressions \eqref{eq: quadRad_hPlus}, \eqref{eq: quadRad_hCross}, and \eqref{eq: quadRad_Phi}, we now examine the effect of eccentricity on gravitational radiation.\\
Figs. \ref{fig: hplus_E}, \ref{fig:hcross_E} and \ref{fig:Phi_E} illustrate the gravitational-wave amplitudes for the three polarization modes and different eccentricities. In all cases, higher eccentricities tend to suppress the amplitudes. Specifically, the lowest eccentricity considered, $e=1.1$ results in the largest amplitudes for both tensor and scalar modes, whereas the highest eccentricity, $e=1.7$, produces the smallest amplitudes across all modes. Notably, the scalar mode wave form also exhibits Bremsstrahlung-type radiation comparable with that seen in the two tensor mode wave forms.\\

This suppression at higher eccentricities can be understood through the relationship between the eccentricity, the impact parameter $b$ and the tangential velocity $v_0$, as described by equation \eqref{ecc}. Indeed, a larger impact parameter corresponds to a more distant hyperbolic encounter, where the gravitational interaction between the two black holes is weaker, leading to minimal acceleration and, consequently, reduced gravitational-wave emission. Similarly, a higher asymptotic velocity implies a shorter interaction duration, limiting the relative acceleration between the black holes. As a result, both larger impact parameters and higher tangential velocities contribute to the observed suppression of gravitational waves at higher eccentricities, consistent with the findings in Figs. \ref{fig: hplus_E}, \ref{fig:hcross_E}, \ref{fig:Phi_E} and in \cite{PhysRevD.108.024003}.

\subsection{Relativistic corrections to close hyperbolic encounters}
\label{sec:pre}
In Sections \ref{sec:nonpre} and \ref{sec:ecc} we studied the Newtonian orbital regime and the effect of varying eccentricity. The gravitational influence of the companion black hole generates an orbital deviation on the incoming object during the encounter. This effect can also be studied through the inclusion of a precession-like parameter into the orbital mechanics equations, Noting that in the purely Newtonian case, $\alpha = 0$.\\
A visualization of this effect can be seen in Figure \ref{fig:KeplerCHEfig}, where a non-zero $\alpha$ causes an additional orbital deviation. This effect is studied through the evolution of the trajectory's radial coordinate. Such a deviation in the radial trajectory has been considered in the literature for the GR scenario \cite{PhysRevD.109.064001, Capozziello_DeLaurentis_DePaolis_Ingrosso_Nucita_2008} 
and is given by
\begin{equation}
\label{r_pr}
    r_{pr}(\varphi)=\frac{a(e^2-1)}{1+e\cos{[(1-\alpha)\varphi]}}\;,
\end{equation}
 where $\alpha$ is the parameter encoding orbital precession, and is related to the other orbital parameters and observable quantities by \cite{PhysRevD.109.064001}\footnote{A subtlety worth noting is that different gravitational potentials will yield different precession relations. There is indeed literature considering modifications to the GR precession caused by gravitational potentials present in alternative theories, a sample of which can be found in \cite{Martino_Lazkoz_Laurentis_2018, Tedesco_Capolupo_Lambiase_2024, DeMartino_dellaMonica_DeLaurentis_2021, PhysRevD.84.024020, d2021s, PhysRevD.85.124004}.
 }.
\begin{equation}\label{prececc}
    \alpha = \frac{3 \textbf{G}^2 \mathcal{M}^2}{c^2 L^2} = \frac{3 R_s}{2(e+1)r_{min}}\;,
\end{equation}
where $L =b \, v_0$ is the angular momentum per unit mass, $R_s$ is the Schwarzschild radius, and $\mathcal{M}$ is the mass of the central black hole. In the latter expression \eqref{prececc} we have already included the $f(R)$ modification to the gravitational constant. \\
Once the precession has been taken into account, we now calculate the quadrupole moment and the trace of the quadrupole moment as
\begin{align}
    Q^{ij}_{\;\;pr} &= \mu r_{pr}^2(\varphi)
    \begin{pmatrix}
        \cos^2{\varphi} & \cos{\varphi}\sin{\varphi} & 0 \\
        \cos{\varphi}\sin{\varphi} & \sin^2{\varphi} & 0 \\
        0 & 0 & 0\\
    \end{pmatrix} \;,\label{Qij_pr}\\
    \quad \text{and} \quad \nonumber\\
    Q_{pr} &= \frac{2 e \mu v_{0}^2(1+\alpha)^2}{-1+e^2} \left\{ e+\cos{[(-1+\alpha)\varphi]} \right\} \label{Q_pr}\;.
\end{align}

Consequently, according to Eqs. \eqref{h_plus_fR}, \eqref{h_cross_fR} and \eqref{Phi_fR} the quadrupole radiation expressions for the three polarization modes with precession ($pr$) are given by
\begin{eqnarray} 
\label{h_Plus}
        \tilde{h}_{+\,pr} &=&   \left[ \frac{2 \textbf{G} \mu  v_{0}^2}{(e^2-1)\mathcal{R}c^4} \right]  \Bigg((\alpha e^2(\alpha -2)-2) \cos (2 \varphi )\nonumber\\
        &+& e \{(\alpha ^2-5) \cos ((\alpha +1) \varphi )\nonumber\\
        &+&[(\alpha -4) \alpha -1] \cos ((\alpha -3) \varphi )\nonumber\\
        &-&\alpha  e \cos (2 (\alpha -2) \varphi )+(\alpha -2) e \cos (2 \alpha\varphi)\}\Bigg)\,,\nonumber\\
        &&
\end{eqnarray}
\begin{eqnarray}
\label{h_Cross}
        \tilde{h}_{\times\,pr} &=& \left[ \frac{\textbf{G} \mu v_{0}^2}{(e^2-1)\mathcal{R}c^4 } \right] \bigg( 2 [(\alpha -2) \alpha  e^2-2] \sin (2 \varphi)\nonumber\\
        &+&e [(\alpha ^2-5) \sin ((\alpha +1) \varphi )\nonumber\\
        &+&( -\alpha^2 +4\alpha+1) \sin ((\alpha -3) \varphi )\nonumber\\
        &+&\alpha  e \sin (2 (\alpha -2) \varphi )+(\alpha -2) e \sin (2 \alpha  \varphi )]\bigg)\;,\nonumber\\
        &&
\end{eqnarray}
and 
\begin{multline}  \label{phiPrecession}
    \Phi_{pr} = \frac{4\textbf{G}\,{\rm e}^{-m_\text{S}\mathcal{R}}}{\mathcal{R}c^2} \\
    \times \left\{ M + \frac{e\mu v_0^2(\alpha-1)^2 [\cos{((\alpha-1)\varphi)}+e]}{c^2(e^2-1)} \right\}\;.
\end{multline}   

\begin{figure}
    \centering
    \includegraphics[width = 1.05\linewidth]{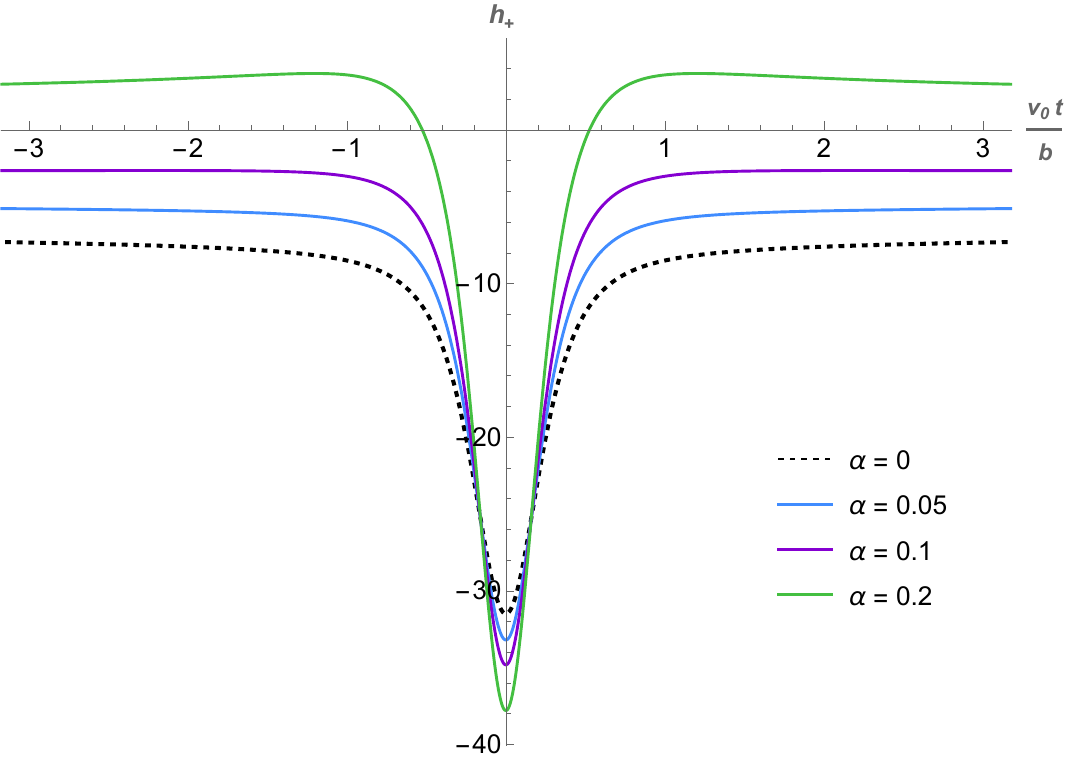}
    \caption{$h_{+\,pr}$ strain with varying precession, eccentricity set to 1.2. Strains are scaled by the constant prefactor $\frac{\text{\bf G}\mu v_0^{2}}{\mathcal{R}c^4}$.}
    \label{fig:hplusPRE}
\end{figure}

\begin{figure}
    \centering
    \includegraphics[width = 1.08\linewidth]{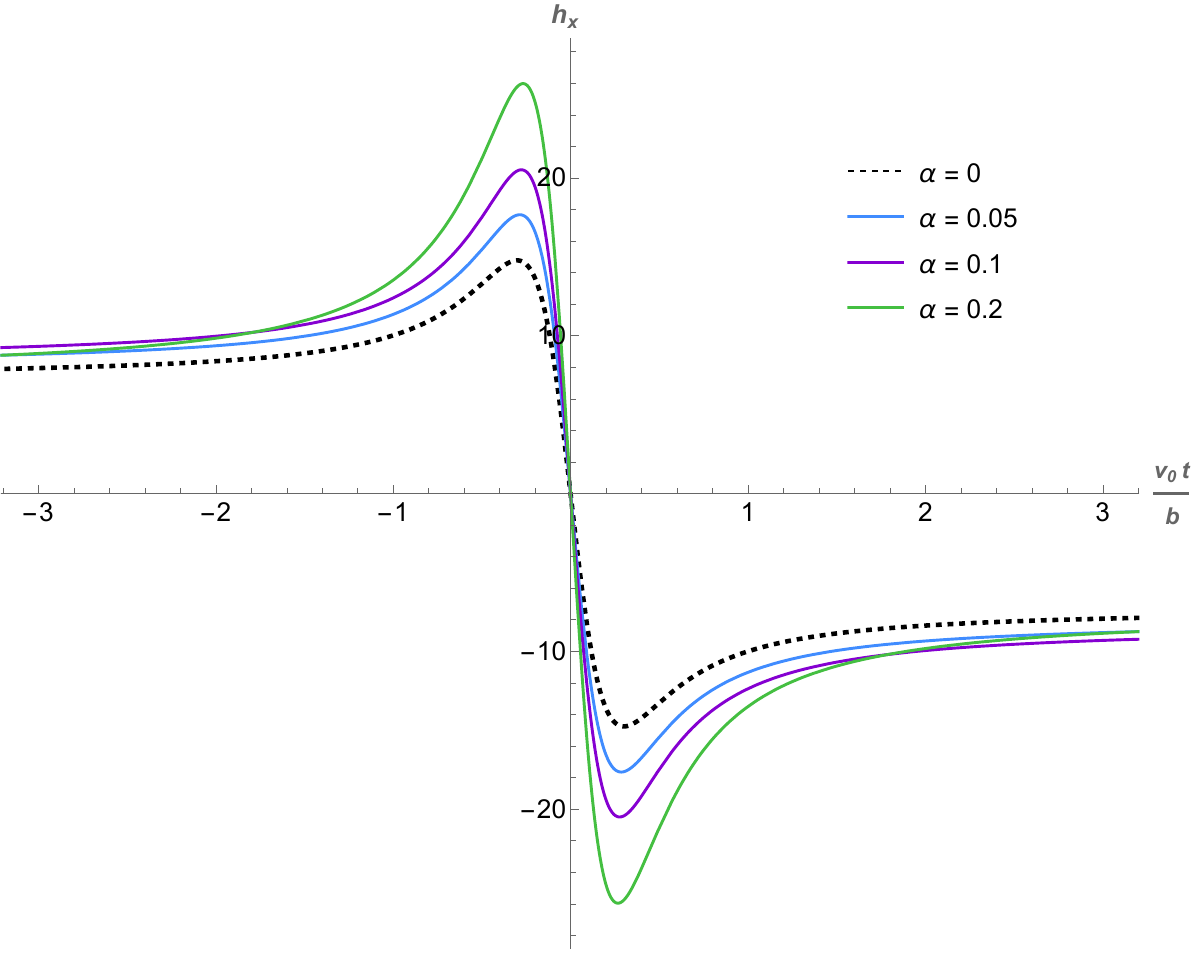}
    \caption{$h_{\times\,pr}$ strain with varying precession, eccentricity set to 1.2. Strains are scaled by the constant prefactor $\frac{\text{\bf G}\mu v_0^{2}}{\mathcal{R}c^4}$.}
    \label{fig:hcrossPRE}
\end{figure}

\begin{figure}
    \centering
    \includegraphics[width = 1.065\linewidth]{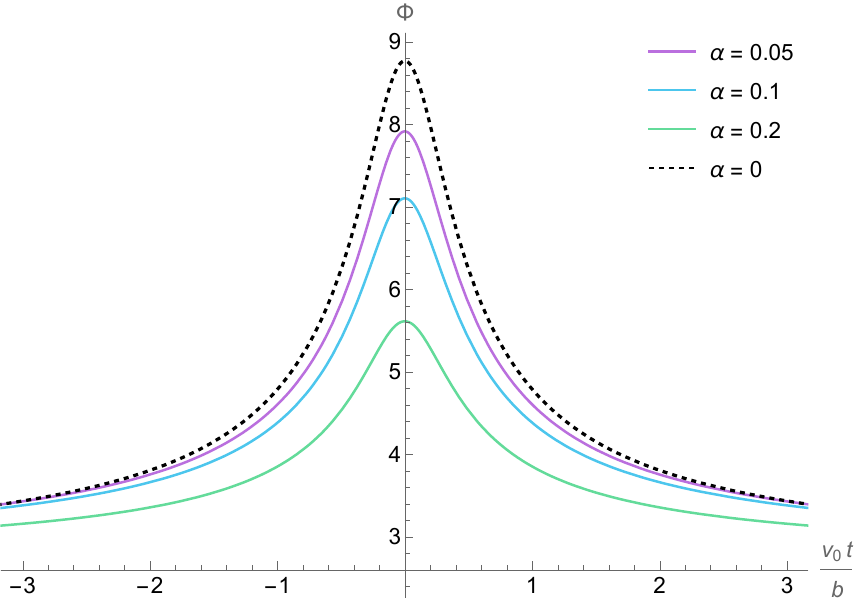}
    \caption{Scalar strain $\Phi_{pr}$ with varying precession, eccentricity set to 1.2 and $m_S = 1\times10^{-33} \,\,\, \text{eV}/c^2$. Strains are scaled by the constant prefactor $\frac{4\textbf{G}}{\mathcal{R}c^2}$. The strain amplitude is suppressed with higher $\alpha$ values, contrary to the behavior of the tensor mode amplitudes.}
    \label{fig:phiPRE}
\end{figure}

These contributions to the quadrupole radiation are plotted in Figs. \ref{fig:hplusPRE}, \ref{fig:hcrossPRE} and \ref{fig:phiPRE} respectively. On the one hand, Figs. \ref{fig:hplusPRE} and \ref{fig:hcrossPRE} indicate that, for the tensor radiation modes, higher precession values yield greater gravitational strain.\\

The behavior of the $\tilde{h}_{+\,pr}$ and $\tilde{h}_{\times\,pr}$ modes is explained by the relationship between precession and eccentricity, as seen in \eqref{prececc}, showing that lower values of precession lead to higher values of eccentricity, and thus greater suppression of the gravitational strain produced during the hyperbolic encounter. Conversely, higher values of precession lead to lower eccentricities and thus, greater resulting gravitational strain.\\

On the other hand, the scalar mode $\Phi_{pr}$ shows sup\-pressed strain for higher $\alpha$ values as seem in Fig. \ref{fig:phiPRE}. This can be explained by the $(\alpha-1)^2$ pre-factor in \eqref{phiPrecession}, which mo\-du\-lates the amplitude of the wave. While  values of $\alpha > 1$ are mathematically admissible, valid precession va\-lues are constrained by the ratio of the Schwarzschild radius to the periastron distance, i.e:   $R_s \neq r_{min}$, along with $\alpha \ge 0$ in the current model\footnote{$\alpha < 0$ would imply a change in the orbital direction \cite{Laurentis_Martino_Lazkoz_2018}.}. Additionally, the hyperbolic orbit fixes $e>1$ and from \eqref{prececc}, we see that $\alpha < \dfrac{3}{4}$. Thus, the only valid precession values cause the $(\alpha-1)^2$ pre-factor cause an overall decrease in the scalar mode amplitude, and higher precession values cause greater suppression in $\Phi_{pr}$. A Taylor expansion of $\Phi_{pr}$ in \eqref{phiPrecession} around $\alpha = \frac{3}{4}$, given by 
\begin{align}
\label{PhiTaylorExp}
    \Phi_{pr} \approx& \Big({\rm e}^{-m_{\rm S} {\mathcal R}} \Big)\left(c^2 M+\frac{e \mu  v_{0}^2 \left(e+\cos \left(\frac{\varphi}{4}\right)\right)}{16 \left(e^2-1\right)}\right)\nonumber
    \\&+\frac{\left(\alpha -\frac{3}{4}\right) e \mu  v_{0}^2} {4  \left(e^2-1\right)} \bigg[\varphi \sin\left(\frac{\varphi}{4}\right)-8 \left(e+\cos \left(\frac{\varphi}{4}\right)\right) \nonumber
    \\&+\frac{\left(\alpha -\frac{3}{4}\right)}{2} \bigg(32 e-16 \varphi \sin \left(\frac{\varphi}{4}\right) \nonumber
    \\&-\left(\varphi^2-32\right) \cos \left(\frac{\varphi}{4}\right)\bigg)\bigg] +{\mathcal O}\left(\left(\alpha -\frac{3}{4}\right)^3\right)
\end{align}
yields a concave function for all values of $\alpha < \dfrac{3}{4}$ when all other parameters are held constant, visually presented in fig \ref{fig:TaylorExpPlot} below.  While the scalar mode amplitude would increase for precession values greater than 1, shown in the grey shaded region of fig \ref{fig:TaylorExpPlot}, values of $\alpha>\frac{3}{4}$ cannot be considered due to the reasoning discussed above.
\begin{figure}[h!]
    \centering
    \includegraphics[width=1.05\linewidth]{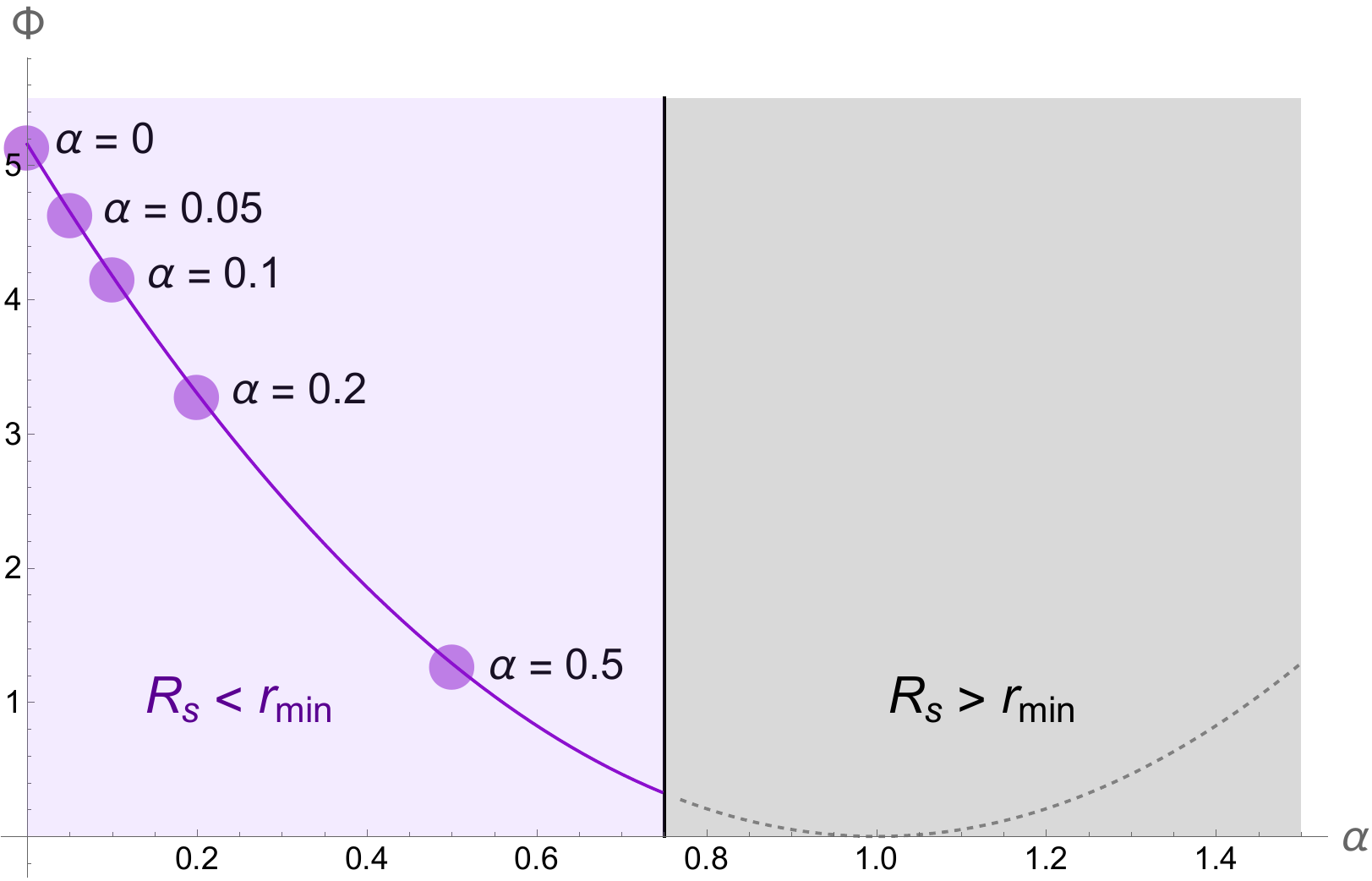}
    \caption{Taylor Expansion of $\Phi_{pr}$ as per \eqref{PhiTaylorExp}
    around $\alpha = \frac{3}{4}$ against $\alpha \in [\,0\,, 0.75\,)$, corresponding to physically valid regions of space shaded light purple, and $\alpha \in (\, 0.75, \, 1.5\,)$ for nonphysical grey-shaded regions inside the Schwarzschild radius of the central object, taken at periastron (i.e: $\varphi = 0$). The solid black line corresponds to $R_S = r_{min}$. Values of $\alpha$ chosen in figs. \ref{fig:hcrossPRE}, \ref{fig:hplusPRE},and \ref{fig:phiPRE} are highlighted on this plot. For illustrative purposes we have considered  $e = 1.5$, $m_{\rm S} = 1 \times 10^{-33} \;\; \text{eV}/c^2$. We see that for all viable precession values, $\Phi$ is an overall and consistently decreasing function, visually explaining the trends seen in fig \ref{fig:phiPRE}.}
    \label{fig:TaylorExpPlot}
\end{figure}
It is worth noting that periastron values close to the Schwarzschild radius would require post-Newtonian corrections and Numerical Relativity analysis me\-thods \cite{PhysRevD.109.064001, PhysRevD.89.081503, Nelson_2019, PhysRevD.108.124016,Vinckers:2023oxa,fontbuté2024numericalrelativitysurrogatemodelhyperbolic}, a scenario which is beyond the scope of this work. As such, precession values $\alpha \gtrsim \frac{3}{4}$ are not considered. Additionally,  we note that values of $\alpha < 10^{-3}$ are admitted but are not considered since their effect on the gravitational-wave strains becomes negligible. A full discussion of the orbital parameters validity range can be found in \cite{PhysRevD.109.064001, Teuscher_Barrau_Martineau_2024, Dandapat_2023}.\\

Finally, the comparison of  the tensor and scalar mode behaviors shows that the tensor strains increase when the periastron distance is closer to the Schwarzschild radius, while the $f(R)$ scalar mode is suppressed. For a sample of works investigating CHEs in post-Newtonian formalism in GR, see \cite{Loutrel_Yunes_2017, bini2024searchhyperbolicencounterscompact, PhysRevD.96.064021, Roskill_Caldarola_Kuroyanagi_Nesseris_2024}, whereas for a massless scalar-tensor theory some results can be found in  \cite{Jain_2023}.
\section{Constrains on Scalar Massive Mode Time Delay}\label{sec: scalarModeMass}
Since CHEs are suitable models for multiple astrophysical environments, they provide ample grounds to constrain $f(R)$ models through observable quantities. One such quantity is the detection time delay between the arrival of the tensor and scalar modes respectively.\\

While it is known that tensor modes arising from GR propagate at the speed of light, $c$ \cite{2Abbott_Abbott_Abbott_Acernese_Ackley_Adams_Adams_Addesso_Adhikari_Adya_etal._2017,Bailes_Berger_Brady_Branchesi_Danzmann_Evans_Holley-Bockelmann_Iyer_Kajita_Katsanevas_etal._2021, Baker_Calcagni_Chen_Fasiello_Lombriser_Martinovic_Pieroni_Sakellariadou_Tasinato_Bertacca_etal._2022, Vitale_2021}, many $f(R)$ theories predict a scalar propagation speed, $c_S$ (such that $c_S < c$, necessarily), from which the mass of the scalar mode, $m_\text{S}$, can be constrained \cite{Dash_Yadav_Verma_2024}. Alternatively, the constrained $m_S$ from any chosen ${f(R)}$ theory can be used to determine the time delay $\Delta t$ occurring between the arrivals of the tensor and scalar modes. Thus providing estimates for an observable quantity that is testable for current and future gravitational wave observatories.\\

We start by defining the propagation speed of a wave packet as the group velocity, and using \eqref{ms_eqn} and  \eqref{ScalarWaveSolution}, we conclude that 
\begin{equation}\label{cs}
    c_\text{S} = \dfrac{\partial \omega}{\partial k} = \dfrac{1}{\cosh{\xi}} \;.
\end{equation}
Consequently, recovering the dimensionality of $c$ and resorting to the dispersion relation of the scalar mode
\begin{equation}
    \omega = \frac{c}{\hbar}\sqrt{\hbar^2 k^2 + c^2 m_\text{S}^2}\;,
\end{equation}
and the $c_\text{S}$ definition as a group velocity in \eqref{cs}, it is straightforward to show
\begin{equation}\label{masscalc}
    m_\text{S} = \frac{\hbar\omega}{c^2}\sqrt{1 - \left(\frac{c_\text{S}}{c}\right)^2} \;,
\end{equation}
or alternatively \cite{PhysRevD.108.024003}
\begin{align}
    m_\text{S} &= 4.14\times 10^{-15}\sqrt{1 - \Big(\dfrac{c_\text{S}}{c}\Big)^2}\left[ \dfrac{f}{\text{Hz}}\right]\left[\frac{\text{eV}}{c^2}\right]\;. 
\end{align}

We now re-frame $m_\text{S}$ in terms of observable quantities.
Due to the difference in propagation speeds, there will be a delay in detection times for the tensor and scalar modes generated at some distance $\mathcal{R}$ from the observer, namely 
\begin{equation}
\label{Delta_t}
  \Delta t = \frac{\mathcal{R}}{c_\text{S}} - \frac{\mathcal{R}}{c} \;,
\end{equation}
which yields the estimated ratio of propagation speeds as 
\be \label{ratio}
\frac{c_\text{S}}{c} = \frac{1}{1+\frac{c \Delta t }{\mathcal{R}}}\;.
\ee

From \eqref{masscalc} and \eqref{ratio}, an estimate for  $m_\text{S}$ in terms of observable quantities is given by
\begin{equation}
\label{ms_final}
    m_\text{S} = 4.14\times 10^{-15}\sqrt{1 - \left(\frac{1}{1+\frac{c \Delta t }{\mathcal{R}}}\right)^2}\left[ \dfrac{f}{\text{Hz}}\right]\left[\frac{\text{eV}}{c^2}\right]\;.
\end{equation}
, calculated by rearranging \eqref{ms_final} as
\begin{equation}
    \Delta t = \frac{\mathcal{R}}{c}\left(\frac{1}{\sqrt{1 - \left(\frac{m_{\rm S}\,[{\rm eV}/c^2]}{4.14\times10^{-15}\,}\right)^2}}-1\right)\;,
\end{equation}
where the frequency is set to 1 Hz, and noting that the scalar mass is given in units of $\left[ \frac{\text{eV}}{c^2}\right]$. Table \ref{table: Final_Deltat_Table} shows a summary of time delay values for various distances and values of $m_S$, the latter obtained from three viable classes of $f(R)$ gravity models - Starobinsky \cite{Starobinsky_2007,Lambiase_2021}, Hu-Sawicki \cite{Parbin_Goswami_2021, Hu_Sawicki_2007} and the Exponential \cite{Appleby_Battye_2008, Nojiri_Odintsov_2011} models respectively. The distances used therein are estimates for sources interacting through  CHEs, occurring in the central galactic region, globular clusters and proposed primordial black-hole populations.\\

We see from Table \ref{table: Final_Deltat_Table} that the value of $m_{\rm S}$ has a significant impact on the observed time delay, noting that the Starobinsky and Hu-Sawicki models provide the best chances of $\Delta t$ detection across all three distance choices. Additionally, astrophysical sources at cosmological distances provide the highest likelihood of detecting this time delay, due to the fact that closer sources have smaller $\Delta t$ values. The exponential $f(R)$ model, relating to a $f(R)$ modification on cosmological scales yields time delays that are undetectably small with current observational tools. It is worth noting that the LIGO detection GW170817 has placed $m_{\rm S}\lesssim 1\times10^{-22} \,\text{eV}/c^2$ \cite{2Abbott_Abbott_Abbott_Acernese_Ackley_Adams_Adams_Addesso_Adhikari_Adya_etal._2017,dejrah2025frgravitygravitationalwaves} as an upper bound, and is respected in the values chosen in Tables \ref{table: Final_Deltat_Table} and \ref{table: Rst_Table}.
\begin{table*}[ht]
\begin{tabular}{c|c|c|c|}
\cline{2-4}
\multicolumn{1}{l|}{}                                                                                                     & \textbf{\begin{tabular}[c]{@{}c@{}}Galactic Centre\\ $\mathcal{R} = 10\,{\rm kpc}$\end{tabular}} & \textbf{\begin{tabular}[c]{@{}c@{}}Globular Cluster \\ $\mathcal{R} = 100\,{\rm kpc}$\end{tabular}} & \textbf{\begin{tabular}[c]{@{}c@{}}Primordial Black Hole\\ Population\\ $\mathcal{R} = 4200\,{\rm Mpc}$\end{tabular}} \\ \hline
\multicolumn{1}{|c|}{\textbf{\begin{tabular}[c]{@{}c@{}}Starobinsky \\ ($m_{\rm S} = 10^{-22}\, {\rm eV}/c^2$)\end{tabular}}}  & $3\times 10^{-4}$s                                                                       & $3\times 10^{-3}$s                                                                          & $1.2\times 10^{2}$s                                                                                           \\ \hline
\multicolumn{1}{|c|}{\textbf{\begin{tabular}[c]{@{}c@{}}Hu-Sawicki \\ ($m_{\rm S} = 10^{-24}\,{\rm eV}/c^2$)\end{tabular}}} & $3\times10^{-8}$s                                                                        & $3\times10^{-7}$s                                                                           & $1.2\times 10^{-2}$s                                                                                          \\ \hline
\multicolumn{1}{|c|}{\textbf{\begin{tabular}[c]{@{}c@{}}Exponential\\ ($m_{\rm S} = 10^{-33}\,{\rm eV}/c^2$)\end{tabular}}}  & $3\times 10^{-26}$s                                                                      & $3 \times 10^{-25}$ s                                                                       & $1.2 \times 10^{-20}$s                                                                                        \\ \hline
\end{tabular}
\caption{$\Delta t$ values (in seconds) for different distances of reference, calculated with $m_{\rm S}$ values as obtained from three viable theories of $f(R)$ gravity.}
\label{table: Final_Deltat_Table}
\end{table*}


\section{Ratio Of scalar to tensor Amplitudes}
\label{sec: STRatio}
The ratio of scalar-to-tensor amplitudes, $\mathcal{R}_{\text{ST}}$, serves as another observable that can be used to test deviations from GR by current and future gravitational-wave detectors. Having determined both the scalar and tensor mode amplitudes in preceding sections, we can now compute $\mathcal{R}_{\text{ST}}$ using 
\begin{table*}[ht]
\begin{tabular}{c|ccc|ccc|ccc|}
\cline{2-10}
\multicolumn{1}{l|}{}                           & \multicolumn{3}{c|}{\begin{tabular}[c]{@{}c@{}}{\textbf{  Galactic centre}}\\ $m_\text{S} = 10^{-22}\text{ eV}/c^2$\end{tabular}} & \multicolumn{3}{c|}{\begin{tabular}[c]{@{}c@{}}{\textbf{Globular clusters}}\\ $m_\text{S} = 10^{-24}\text{ eV}/c^2$\end{tabular}} & \multicolumn{3}{c|}{\begin{tabular}[c]{@{}c@{}}
{\bf Primordial black-hole}\\ {\bf population}\\ $m_\text{S} = 10^{-33}\text{ eV}/c^2$\end{tabular}} \\ \cline{2-10} 
\multicolumn{1}{l|}{}                           & \multicolumn{1}{c|}{\text{{\it e} = 1.1}}       & \multicolumn{1}{c|}{\text{{\it e} = 1.3}}       & \text{{\it e} = 1.7}      & \multicolumn{1}{c|}{\text{{\it e} = 1.1}}        & \multicolumn{1}{c|}{\text{{\it e} = 1.3}}       & \text{{\it e} = 1.7}       & \multicolumn{1}{c|}{\text{{\it e} = 1.1}}             & \multicolumn{1}{c|}{\text{{\it e} = 1.3}}             & \text{{\it e} = 1.7}             \\ \hline
\multicolumn{1}{|c|}{\text{$\alpha = 0$}}     & \multicolumn{1}{c|}{0.6852}                 & \multicolumn{1}{c|}{0.7038}                 & 0.6945                & \multicolumn{1}{c|}{0.6851}                  & \multicolumn{1}{c|}{0.7044}                & 0.6951                 & \multicolumn{1}{c|}{0.6852}                       & \multicolumn{1}{c|}{0.7044}                       & 0.6951                       \\ \hline
\multicolumn{1}{|c|}{\text{$\alpha = 0.01$}} & \multicolumn{1}{c|}{0.6665}                 & \multicolumn{1}{c|}{0.6841}                 & 0.6733                & \multicolumn{1}{c|}{0.6833}                  & \multicolumn{1}{c|}{0.7023}                 & 0.6929                 & \multicolumn{1}{c|}{0.6833}                       & \multicolumn{1}{c|}{0.7024}                       & 0.6929                       \\ \hline
\multicolumn{1}{|c|}{\text{$\alpha = 0.1$}}   & \multicolumn{1}{c|}{0.5179}                 & \multicolumn{1}{c|}{0.5249}                 & 0.5077                & \multicolumn{1}{c|}{0.5179}                 & \multicolumn{1}{c|}{0.5249}                & 0.5077                & \multicolumn{1}{c|}{0.5179}                       & \multicolumn{1}{c|}{0.5249}                       & 0.5077                      \\ \hline
\end{tabular}
\caption{$\mathcal{R}_{\text{ST}}$ for different mass scales $m_\text{S}$, eccentricities and precession values, showing decreasing values with increased $\alpha$ due to suppression of $|\Phi|$. Notably, values reported herein indicate considerable deviation from GR and provide potential observational predictions.}
\label{table: Rst_Table}
\end{table*}
\begin{equation}
    \mathcal{R}_{\text{ST}} = \frac{|\Phi|}{|h_{ij |}}\,,
\end{equation}
which is  calculated by performing the averaged peak-to-peak amplitudes over one angular period, $\varphi \in [ 0, 2\pi )$ using the tensor and scalar mode strains according to equations \eqref{h_Plus}, \eqref{h_Cross} and \eqref{phiPrecession} as follows,
\begin{align}   
    \mathcal{R}_{\text{ST}} = \frac{\big< \ddot Q \big> }{ \big < \ddot h \big > }\;.
\end{align}
Table \ref{table: Rst_Table} reports $\mathcal{R}_{\text{ST}}$ values for a sample of three constant precession values set to 
$\alpha = 0,\,0.01,\,0.1$
against a subset of previously investigated eccentricity values, in particular 
$e=1.1,\,1.3,\, 1.7$. Due  to its dependence on the scalar mode quadrupole strain, $\mathcal{R}_{\text{ST}}$ is implicitly sensitive to the choice of the scalar mode mass, $m_\text{S}$. Thus, for illustrative purposes, Table \ref{table: Rst_Table} con\-si\-ders three values of $m_{\text S}$ which correspond to values obtained from viable $f(R)$ models. 
In particular $m_\text{S} = 10^{-22} \text{ eV}/c^2$ has been applied to both the Galactic centre and intra-galactic globular cluster scales, $m_{\text S} = 10^{-24} \text{ eV}/c^2$ to globular cluster scales, and finally,  $m_{\text S} =  10^{-33}\text{ eV}/c^2$ to primordial black-hole scales.
It is noteworthy that $m_{\text S}$ depends on the gravity theory used \cite{PhysRevD.108.024003}, however since we are considering a fully general $f(R)$ scenario, the values chosen for $m_\text{S}$ may occur in a broad range of $f(R)$ theories - and choices of parameters therein. In this regard, we note that a mass choice compatible with inflationary scales would rapidly suppress the scalar mode and would greatly reduce the possibility of scalar mode radiation detection {\it c.f.} Table \ref{table: Final_Deltat_Table}.\\

Table \ref{table: Rst_Table} reports $\mathcal{R}_{\text{ST}} < 1$ for all combinations of eccentricity, precession and $m_\text{S}$ mass values - indicating a suppression of the scalar amplitude relative to tensor amplitudes. This is an expected result, due to the effective screening of $f(R)$ scalar mode by the Chameleon mechanism \cite{Katsuragawa_Nakamura_Ikeda_Capozziello_2019} and is consistent with current predictions of $\mathcal{R}_{\text{ST}}$ from $f(R)$ models, ranging between  $\sim10^{-3}$ and $\sim10^{-1}$ \cite{Takeda_Manita_Omiya_Tanaka} \footnote{An example of tests used to constrain $\mathcal{R}_{\rm ST}$ includes dipole radiation measurements from GW170817 \cite{Ezquiaga:2017ekz}.}. Notably, while contemporary studies of $\mathcal{R}_{\text{ST}}$ from modified theories primarily consider binary systems, due to the fact that $\mathcal{R}_{\rm ST}$ is likely to be larger for these sources, we find that CHEs can be a novel testing ground for $\mathcal{R}_{\rm ST}$ studies. Table \ref{table: Rst_Table} shows $\mathcal{R}_{\text{ST}}$ values considerably higher than these estimates. This indicates a strong scalar mode contribution and, eventually, a significant deviation from GR. Thus, results from Table \ref{table: Rst_Table} imply that burst gravitational waves in general, and CHEs in particular, could provide a substantial testbed for modified theories.\\

\begin{figure}
    \centering
    \includegraphics[width= 1.05\linewidth, height = 0.55\linewidth]{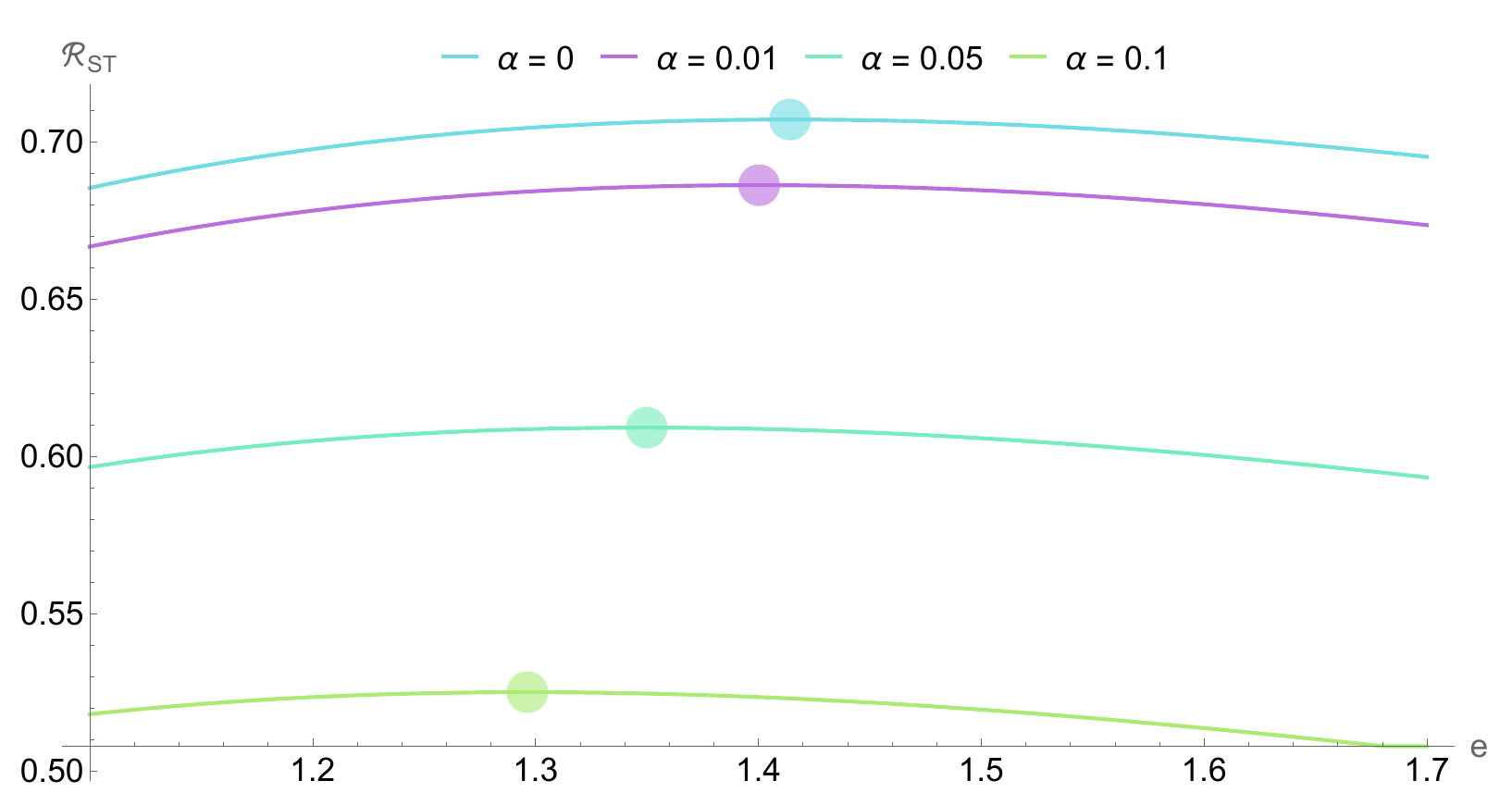}
    \caption{Variation of $\mathcal{R}_{\text{ST}}$ with respect to eccentricity, for different values of orbital precession and all other astrophysical parameters are the same as previous figures.}
    \label{fig:RST_Plot}
\end{figure}

\begin{figure} [ht!]
    \centering
    \includegraphics[width=1.05\linewidth]{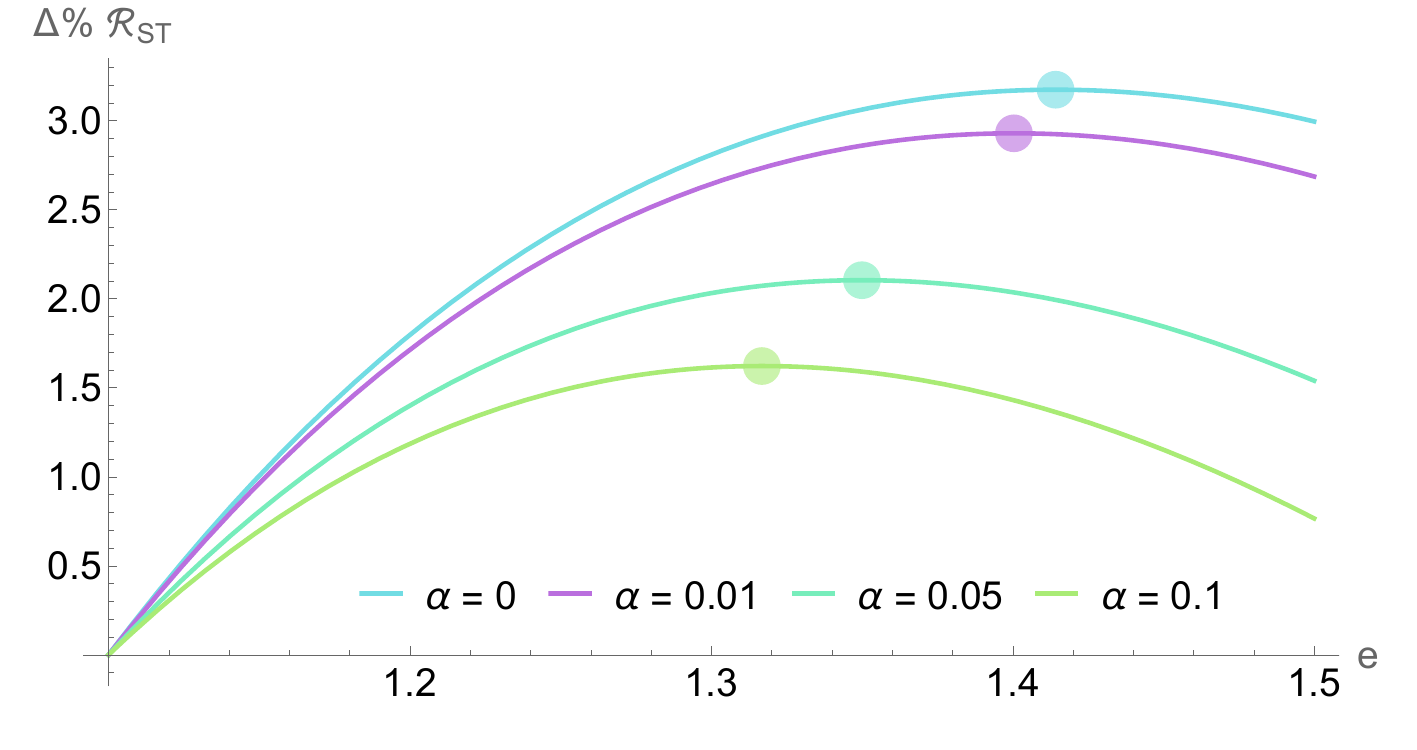}
    \caption{Percentage change of $\mathcal{R}_{\text{ST}}$ against varying values of eccentricity, starting at $e=1.1$, with constant precession. Precession values considered herein are the same as those considered in Section \ref{sec:pre}. Higher $\Delta \% \mathcal{R}_{\text{ST}}$ occurs for lower $\alpha$, and maximizing $e$ values decrease with higher $\alpha$, consistent with results in Table \ref{tab: MaxEccentricity}. Values of $e$ for maximum $\mathcal{R}_{\text{ST}}$ are marked on each plot by a circle.}
\label{fig:PercentChange}
\end{figure}

Additionally, Table \ref{table: Rst_Table} shows $\mathcal{R}_{\text{ST}}$ decreasing with increasing precession, consistent with results discussed in Section \ref{sec:pre}, due to the suppression of the scalar amplitude as $\alpha \rightarrow \frac{3}{4}$. Table \ref{table: Rst_Table} also indicates a variation of the scalar-tensor ratio with respect to eccentricity, for differing fixed values of precession. For each value of $\alpha$, there is a corresponding eccentricity for which maximizes $\mathcal{R}_{\text{ST}}$, above and below which the values of $\mathcal{R}_{\text{ST}}$ decrease. This relationship can be attributed to $i)$ the coupling between eccentricity, precession, as seen in \eqref{h_Plus}, \eqref{h_Cross},  and \eqref{phiPrecession}; $ii)$ the relationship between eccentricity and the tangential velocity and impact parameter according to \eqref{ecc}, and $iii)$ the relation between $\alpha$ and the ratio of the Schwarzschild radius and periastron distance as seen in \eqref{prececc} and discussed in Section \ref{sec:pre}. Fig. \ref{fig:RST_Plot} visualizes the eccentricity dependence of $\mathcal{R}_{\text{ST}}$ for each precession value. We see that while the variation is small over the range of eccentricities chosen,  there is an eccentricity value for each precession that yields a maximal $\mathcal{R}_{\text{ST}}$ value, the latter reported in Table \ref{tab: MaxEccentricity}. Table \ref{tab: MaxEccentricity} additionally shows that eccentricity values capable of  maximizing $\mathcal{R}_{\text{ST}}$  decrease with increasing precession. Fig. \ref{fig:PercentChange} provides an illustrative visual representation of this eccentricity  dependence, showing the percentage change of $\mathcal{R}_{ST}$ for each precession value considered in Fig. \ref{fig:RST_Plot}, with the maximum values for each precession corresponding with the eccentricity value generating the highest $\mathcal{R}_{ST}$ for corresponding $\alpha$ value, calculated by 
\begin{eqnarray}
\Delta\%\mathcal{R}_{\text{ST}}(e)=\frac{\mathcal{R}_{\text{ST}, \alpha}(e)-\mathcal{R}_{\text{ST}, \alpha}(e_{\text{min}})}{\mathcal{R}_{\text{ST}, \alpha}(e_{\text{min}})} \times100\;,  
\end{eqnarray}
Where $e \in [\,1.1\,, 1.4\,]$, a subset of the eccentricity values chosen previously. We see that eccentricity values near 1.4 generate the maximum $\mathcal{R}_{\text{ST}}$ for the widest range of precession values, implying that astrophysical systems with orbital eccentricities of this order ($e\sim1.4$) are likely to produce scalar waves that have a higher likelihood of detectability.\\ 

This result suggests that scalar-mode gravitational radiation produced by gravity theories introducing modifications at the dark energy scale (discussed further in \cite{PhysRevD.108.024003}) are likely to be detectable by future gravitational-wave observatories with sensitivity in low frequency bands. Additionally, our results suggest that close hyperbolic encounters are likely to generate detectable $f(R)$ scalar mode radiation for all combinations of parameter values considered herein.

\begin{table}[ht]
    \centering
    \begin{tabular}{|c|c|c|}
        \hline
        Precession $\alpha$ & Maximum $\mathcal{R}_{\text{ST}}$ & Maximizing eccentricity $e$ \\ \hline
        0& 0.7069 & 1.4142\\ \hline
        0.01 & 0.6861 & 1.4004\\ \hline
        0.05 & 0.6090 & 1.3499 \\ \hline
        0.1 & 0.5249 & 1.2964 \\ \hline
    \end{tabular}
     \caption{ Eccentricities yielding maximal $\mathcal{R}_{\text{ST}}$ for each fixed precession value. Results show that the $e$ values maxi\-mizing $\mathcal{R}_{\text{ST}}$  decrease with increased precession  $\alpha$.} 

    \label{tab: MaxEccentricity}
\end{table}

 \section{Conclusions}
 \label{Conclusions}
In the context of $f(R)$ metric theories, this work investigates the contributions of both scalar and tensor modes to the gravitational radiation emitted during close hyperbolic encounters of black holes. These scenarios may span a variety of astrophysical environments containing populations of compact objects, including globular clusters, galactic centers and primordial black-hole populations.\\

We first analyzed these encounters while neglecting orbital precession and subsequently extended the study by incorporating a precession parameter, $\alpha$. In both cases, we included the scalar longitudinal mode characteristic of $f(R)$ gravity, which is absent in General Relativity. Our results confirm the occurrence of bremsstrahlung-type radiation for all polarization modes -- $\tilde{h}_+$, $\tilde{h}_{\times}$, and $\Phi$ -- in both precessing and non-precessing scenarios. Due to the dependence of orbital parameters on incoming tangential velocity and impact parameter, higher eccentricities suppress both tensor and scalar modes. Similarly, high tangential velocities and large impact parameters reduce the interaction between the compact objects, leading to decreased acceleration and, consequently, lower gravitational-wave amplitudes.\\

Furthermore, we demonstrate that the scalar mode radiation is increasingly suppressed for higher precession values, while tensor amplitudes are increased. This suppression arises from a coefficient in the quadrupole radiation formula specific to the scalar mode, (see Eq. \eqref{phiPrecession}), which is absent in the tensor mode expressions, in particular \eqref{h_Plus} and \eqref{h_Cross}. Despite this suppression, the scalar mode amplitude remains comparable to its tensor counterparts. Given that scalar-mode waves generally exhibit lower amplitudes in many gravitational-wave sources \cite{PhysRevD.108.024003}, our findings suggest that close hyperbolic encounters could provide a viable channel for detecting scalar modes in gravitational wave observations.\\

To further explore observational prospects, we examined the time delay predicted by the theoretical scalar mode masses  arising various $f(R)$ theories. We find that scalar modes are significantly suppressed for sources at cosmological distances, while detectable time delays are possible for closer sources. Additionally, we computed the scalar-to-tensor radiation ratio, finding that it is maximized at intermediate eccentricities ($e \sim 1.4$) and low precession values, while minimized for high eccentricities and precession values. This behavior is attributed to the increasing suppression of the scalar mode at higher precession values. The dependence of the quadrupole radiation formulae on orbital parameters also contributes to variations in the scalar-tensor ratio. Remarkably, the ratio remains of order unity for all parameter combinations considered, indicating that $f(R)$ scalar-mode gravitational waves, expected to play a significant role at dark energy scales, may be detectable by future observatories such as LISA, ASTROD-GW, the Einstein Telescope and the Cosmic Explorer 
 \cite{LouisYang_2011, PhysRevD.99.124050, Punturo_Abernathy_Acernese_Allen_Andersson_Arun_Barone_Barr_Barsuglia_Beker_etal._2010, Reitze_Adhikari_Ballmer_Barish_Barsotti_Billingsley_Brown_Chen_Coyne_Eisenstein_etal._2019}.\\
 
To the best of our knowledge, this work represents a first investigation into the detectability of gravitational-wave signals from $f(R)$ theories in hyperbolic encounters of compact objects. However, further theoretical development is required to refine our understanding of these phenomena. Future work should include power and energy radiation calculations, along with a full power spectrum analysis in both frequency and time domains. Additionally, the gravitational memory effect associated with the scalar breathing mode emitted during close hyperbolic encounters warrants further study \cite{Favata_2010, PhysRevD.109.064001, Dandapat_Susobhanan_Dey_Gopakumar_Baker_Jetzer_2024, Du_Nishizawa_2016, Heisenberg_Yunes_Zosso_2023}.
A broader perspective on alternative gravitational radiation scenarios and constraints on $f(R)$ models can be found in \cite{PhysRevD.108.024003, Spherical1, Laurentis_Martino_2022, Narang_Mohanty_Jana_2023, Kalita_Mukhopadhyay_2021, zhou2024scalarinducedgravitationalwaves, Seraj_2021, PhysRevD.108.024010, Achour_Gorji_Roussille_2024}.
Finally, further developments should also explore cases where impact parameters permit multiple orbital passages and release scenarios and broader parameter ranges supported by Numerical Relativity techniques to provide a more precise description of burst gravitational radiation from scattering events in compact object populations.\\

In conclusion, gravitational-wave bursts from close hyperbolic encounters provide a promising avenue for probing astrophysical sources beyond well-studied compact binaries. Moreover, they offer a potential observational window into scalar modes predicted by some paradigmatic extended theories of gravity.\\

\acknowledgments
The authors thank Kelly MacDevette, Adri\'an Casado-Turri\'on and Muzikayise Sikhonde for helpful discussions.
AdlCD acknowledges support from BG20/00236 action (MCINU, Spain), NRF Grant CSUR23042798041, CSIC Grant COOPB23096, Project SA097P24 funded by Junta
de Castilla y Le\'on (Spain) and Grant PID2021-122938NB-I00 funded by MCIN/AEI/10.13039/501100011033 and by {\it ERDF A way of making Europe}.
\bibliography{bib.bib}
\end{document}